\documentclass[12pt,a4paper]{article}
\pdfoutput=1

\usepackage{amsmath,amssymb,amsfonts,mathrsfs,mathtools,graphicx,bbm}
\usepackage{fixltx2e}
\usepackage[font=small]{caption}
\usepackage{lmodern}
\usepackage{booktabs}
\usepackage{dsfont}
\usepackage{slashed}
\usepackage{braket}
\usepackage{hyperref}
\usepackage{xspace}

\usepackage{epsfig}
\usepackage{relsize}
\input epsf
\usepackage{enumitem}
\usepackage{color}

\newcommand{\be}{\begin{equation}}
\newcommand{\ee}{\end{equation}}
\newcommand{\bea}{\begin{eqnarray}}
\newcommand{\eea}{\end{eqnarray}}
\newcommand{\bee}{\begin{enumerate}}
\newcommand{\eee}{\end{enumerate}}
\newcommand{\bei}{\begin{itemize}}
\newcommand{\eei}{\end{itemize}}
\newcommand{\bal}{\begin{equation}\begin{aligned}}
\newcommand{\eal}{\end{aligned}\end{equation}}
\newcommand{\bem}{\left (\begin{matrix}}
\newcommand{\eem}{\end{matrix} \right )}

\newcommand{\la}{\label}

\newcommand{\dn}{\mathop{\mathrm{dn}}\nolimits}

\newcommand{\alg}[1]{\mathfrak{#1}}
\newcommand{\su}{\alg{su}}

\newcommand{\un}{\alg{u}}
\def\sl2{\alg{sl}(2)}

\newcommand{\ads}{${\rm  AdS}_5\times {\rm S}^5\ $}

\def\GN{Gross-Neveu }

\numberwithin{equation}{section}
\makeatletter
 \let\old@startsection=\@startsection
 \let\oldl@section=\l@section
 \renewcommand{\@startsection}[6]{\old@startsection{#1}{#2}{#3}{#4}{#5}{#6\mathversion{bold}}}
 \renewcommand{\l@section}[2]{\oldl@section{\mathversion{bold}#1}{#2}}
\makeatother
\DeclareMathOperator{\Tr}{Tr}

\newcommand{\AdS}{\text{AdS}}

\definecolor{grey}{rgb}{0.4,0.4,0.5}
\definecolor{darkgreen}{rgb}{0,0.5,0}
\definecolor{darkred}{rgb}{0.6,0.0,0}
\definecolor{lightbrown}{rgb}{1,0.9,0.8}
\definecolor{brown}{rgb}{0.6,0.3,0.3}
\definecolor{darkblue}{rgb}{0,0,0.5}
\definecolor{darkmagenta}{rgb}{0.5,0,0.5}

\definecolor{TCDblue}{rgb}{0.1,0.3,0.6}

\def\pa {\partial}
\def\ov{\over}
\def\bra{\langle}
\def\ket{\rangle}

\def\a {\alpha}
\def\b {\beta}
\def\g {\gamma}
\def\G {\Gamma}

\def\de {\delta}

\def\e{\epsilon}
\def\eps{\epsilon}

\def\L{\Lambda}

\def\vk{\varkappa}
\def\m{\mu}
\def\r {\rho}

\def\s {\sigma}

\def\p{\phi}
\def\P{\Phi}
\def\vp{\varphi}

\def\t{\theta}

\def\u\upsilon
\def\U\Upsilon
\def\vU\varUpsilon
\def\vPi{\varPi}

\def\cA{{\cal A}}

\def\cI{{\cal I}}
\def\cJ{{\cal J}}
\def\cK{{\cal K}}

\def\cN{{\cal N}}
\def\cO{{\cal O}}
\def\cP{{\cal P}}

\def\cZ{{\cal Z}}

\def\bA{{\mathbb A}}

\def\bI{{\mathbb I}}
\def\bJ{{\mathbb J}}
\def\bK{{\mathbb K}}
\def\bL{{\mathbb L}}

\def\bP{{\mathbb P}}

\def\bR{{\mathbb R}}
\def\bS{{\mathbb S}}

\def\bZ{{\mathbb Z}}

\def\da{{\dot a}}
\def\db{{\dot b}}

\def\dk{{\dot k}}
\def\dl{{\dot l}}
\def\dm{{\dot m}}
\def\dn{{\dot n}}

\def\dq{{\dot q}}
\def\dr{{\dot r}}
\def\ds{{\dot s}}

\def\dw{{\dot w}}

\def\dphi{{\dot \p}}
\def\dchi{{\dot \chi}}

\def\dI{{\dot I}}

\def\dK{{\dot K}}
\def\dL{{\dot L}}

\def\dN{{\dot N}}

\def\dR{{\dot R}}

\def\dV{{\dot V}}

\def\x'{\mathaccent 19 x}
\def\y'{\mathaccent 19 y}
\def\n'{\mathaccent 19 n}
\def\u'{\mathaccent 19 u}

\def\et'{\mathaccent 19 \eta}
\def\th'{\mathaccent 19 \theta}
\def\lam'{\mathaccent 19 \lambda}
\def\varet'{\mathaccent 19 \vartheta}
\def\rh'{\mathaccent 19 \rho}
\def\ph'{\mathaccent 19 \phi}
\def\xb'{\mathaccent 19 {\bar{x}}}

\def\d1{{\dot{1}}}
\def\dt{{\dot{2}}}
\def\dth{{\dot{3}}}
\def\df{{\dot{4}}}

\def\lok{\chi_k^-}

\def\sh{\sinh}

\def\sun2{SU(N)$\times$SU(N)}

\begin{document}

\null\vskip-40pt
 \vskip-5pt \hfill
\vskip-5pt \hfill {\tt\footnotesize
TCD-MATH-17-10}
% \vskip-5pt \hfill {\tt\footnotesize HMI-11-05}

\vskip 1cm \vskip0.2truecm
\begin{center}
\begin{center}
\vskip 0.8truecm {\Large\bf Free field representation of the ZF algebra 

\vspace{0.3cm}

of the \sun2 PCF model 
}
\end{center}

\renewcommand{\thefootnote}{\fnsymbol{footnote}}

\vskip 0.9truecm
Sergey Frolov\footnote[1]{ Correspondent fellow at
Steklov Mathematical Institute, Moscow.}\footnote[2]{email: 
frolovs@maths.tcd.ie}
 \\
\vskip 0.5cm

{\it School of Mathematics and Hamilton Mathematics Institute, \\
Trinity College, Dublin 2,
Ireland}

\end{center}
\vskip 1cm \noindent\centerline{\bf Abstract} \vskip 0.2cm 

A free field representation  of the Zamolodchikov-Faddeev algebra of the \sun2 Principal Chiral Field model is constructed, and used to derive an integral representation for  form factors of a multi-parameter family of exponential fields. 

{\it This paper is a tribute to the memory of Prof. Petr Kulish.}

\flushbottom

\newpage

\tableofcontents

\renewcommand{\thefootnote}{\arabic{footnote}}
\setcounter{footnote}{0}

\section{Introduction}

Form factors of a two-dimensional quantum field theory model are matrix elements between the vacuum, $|vac \rangle$,  and $in$-states
\be\la{FFdef}
F_{K_1\ldots K_n}(\t_1,\ldots,\t_n)=\langle vac|O(0)|\t_1,\ldots,\t_n\rangle^{(in)}_{K_1\ldots K_n}\,,
\ee
where $K_i$ is a flavour index of the $i$-th particle, and $\t_i$ is its  rapidity variable related to its energy and momentum by $E_i = m_i\cosh \t_i\,,\ p_i = m_i\sinh \t_i$. 
In a crossing invariant theory  one can express a generic matrix element, $\langle out|O({x})|in \rangle$, of  a local operator $O$ between $in$- and $out$-states
 in terms of the analytically continued form factors.
 
\medskip

The form factors of  integrable two-dimensional relativistic models satisfy certain axioms \cite{KW78}-\cite{Smirnov92}
which have been solved for some models, see e.g. \cite{Smirnov92}-\cite{Babujian2015}. Finding a solution to the axioms is highly nontrivial and requires an extensive use of  the form factors' analytic properties. 
An important observation by Lukyanov \cite{Lukyanov}  (following the ideas in \cite{Lukyanov:1992sc}) is that  for a given model   form factors can be found by constructing a free field representation of its Zamolodchikov-Faddeev (ZF) algebra \cite{Zam,Fad}. Lukyanov's approach has been successfully applied to several models \cite{Lukyanov}, \cite{Lukyanov:1997bp}-\cite{BF13},  in particular in  \cite{Kojima,BF13} to the SU(N) Gross-Neveu (GN) model \cite{GN}. 
An advantage of this approach is that the analytic properties 
 of form factors follow from a free field representation. This might be helpful for nonrelativistic but crossing invariant models where analytic properties of form factors are not completely understood. 
An important example of such a model is provided by the  \ads superstring sigma model in the light-cone gauge \cite{AFrev}. Even though most of the form factors axioms can be generalised to the case \cite{KM}\footnote{Notice that the axioms for a cubic light-cone string field theory vertex   \cite{BJ2015} are very similar to the form factors axioms. }, no solution has been found yet because the analytic properties of  the \ads form factors are  unknown. One may hope that Lukyanov's approach might be more efficient in the \ads case.

\medskip

Another complication of the \ads model is that its symmetry algebra is a sum of two algebras (which are in addition centrally-extended super Lie algebras sharing a central element). 
The only model of such a type for which a free field representation has been constructed is the two-parameter family of integrable models (the SS model) \cite{Fateev:1996}. This representation was found by Fateev and Lashkevich  \cite{FL}. A generalisation of their results to other models is not straightforward. 

\medskip

The goal of this paper is to extend Lukyanov's approach to the \sun2 Principal Chiral Field (PCF) model. The symmetry algebra of the model is obviously $\su(N)\oplus\su(N)$.  
The  ``elementary'' particles of the model transform  in the rank-1 bi-fundamental representation of  \sun2, and they form $r$-particle bound states transforming in the rank-$r$ bi-fundamental representations of  \sun2 \cite{Wiegmann}. 
The exact S-matrix of the PCF model up to a CDD factor \cite{Castillejo:1955ed} is a direct product of the S-matrices of the chiral GN model \cite{Berg:1977dp}-\cite{Koberle:1979ne}, and can be found from the usual requirements of unitarity, analyticity,
crossing symmetry  and the bound state structure \cite{Wiegmann}.
Similarly to the chiral GN model \cite{Kurak:1978su}, anti-particles of  elementary particles are  bound states of $N-1$ elementary particles, and in general
anti-particles of  rank-$r$ particles are rank-$(N$-$r)$ particles. 

\medskip

Nothing is known about form factors of the PCF model for finite $N\ge 3$ except 
the two-particle form factor of the current operator found in \cite{Cubero:2012xi}. 
At infinite $N$ multi-particle form factors of  the renormalised field operator 
were found in 
\cite{Orland:2011rd}, and 
those of 
the current and  energy-momentum tensor operators  in \cite{Cubero:2013iga,Cubero:2014xwa}.  Since up to a twist the  SU(2)$\times$SU(2)$\,=\,$O(4) model can be obtained from the SS model in a special limit $p_1,p_2\to\infty$, much more is known about the $N=2$ case. The form factors of the SS model for a large class of local operators which includes the $U(1)$ currents and energy-momentum tensor  where determined in \cite{Smirnov93} by solving the form factor axioms, and  those for a 3-parameter family of exponential fields in \cite{Fateev:1996} by constructing a free field representation of the ZF algebra. 

\medskip

In this paper a free field representation of the ZF algebra of the \sun2 PCF model  for elementary particles and their bound states  is constructed. It can be used to derive an integral representation for  form factors of a multi-parameter family of exponential fields through Lukyanov's trace formula \cite{Lukyanov}. The determination of the precise form of the exponential fields and their relation to the fields which appear in the Lagrangian of the PCF model is outside the scope of the paper.
For $N=2$ the free field representation and the integral representation should be equivalent 
to those found by Fateev and Lashkevich  \cite{FL}. However, the O(4) limit of their representation is subtle, and in this paper a proof of the equivalence is not attempted. 

\medskip

The outline of the paper is  as follows. In section \ref{gener}  the properties of the scattering matrix, and the particles content of the PCF model are reviewed. Here,  the general idea of the free field representation approach to form factors is also explained. In section \ref{free}  the construction of a representation of the extended ZF algebra and angular Hamiltonian for the PCF model is considered, Green's functions and relations between free fields are listed, and the main properties of the representation are discussed. Next, it is proven in section \ref{zfalg} that the vertex operators constructed in section \ref{free} indeed satisfy the ZF algebra relations.
Then, the derivation of the highest weight bound state vertex operators and bound states is performed in section \ref{boundstates}. 
Finally,  form factors are discussed in section \ref{form}, where general formulae for constructing form factors are established. 
In several appendices the necessary functions are collected and the derivations of some results stated in the main text are presented.

\section{Generalities}\la{gener}

\subsection{The S-matrix of the model}

The spectrum of particles of the  \sun2 PCF model consists of  elementary particles of mass $m$ transforming   in the rank-1 bi-fundamental representation of  \sun2, and their $r$-particle bound states of mass $m_r=m \sin {\pi r\ov N}/\sin {\pi\ov N}$ transforming  in the rank $r=2,\ldots,N-1$  bi-fundamental representation of  \sun2. A rank-$r$ particle with rapidity $\t$ is created by a ZF operator ${\cal A}^\dagger_{K\dK}(\t)$, and annihilated by $\cA^{K\dK}(\t)$ where ${K} = (k_1,\ldots, k_r)$ and ${\dK} = (\dk_1,\ldots, \dk_r)$ have integer-valued components ordered as $1\le k_1< k_2<\cdots< k_r\le N$, and $\d1\le \dk_1< \dk_2<\cdots< \dk_r\le \dN$. The left and right SU(N) groups act on the undotted and dotted indices respectively. 

The S-matrix  for the elementary particles of the PCF model is \cite{Wiegmann}
\begin{equation}  \label{S-matrix}       
\bS^{\rm PCF}(\t) =\chi_{_{CDD}}(\theta)\cdot S(\theta) {\bR(\theta)}
\otimes S(\theta)  {\bR(\theta)}=S_{PCF}(\theta)\, {\bR(\theta)}
\otimes  {\bR(\theta)}\,,  
\end{equation}        
\begin{equation}        
 S(\theta)=       
-\frac{\Gamma \left( \frac{i \theta}{2\pi} \right)       
\Gamma \left(\frac{1}{N}- \frac{i \theta}{2\pi} \right)}       
{\Gamma \left(-\frac{i \theta}{2\pi} \right)       
\Gamma \left(\frac{1}{N}+ \frac{i \theta}{2\pi} \right)} \, , \quad   
\chi_{_{CDD}}(\theta)=\frac{\sinh \left(\frac{\theta}{2}+\frac{i\pi}{N}\right)}{\sinh \left(\frac{\theta}{2}-\frac{i\pi}{N}\right)}   \,.
\end{equation}   
Here  the standard SU(N)-invariant R-matrix is 
\bal       
\bR(\theta)=\frac{\theta\, \bI-\frac{2\pi i}{N}\,\bP} {\theta-\frac{2\pi i}{N}}=\bP_s+ \frac{\theta +\frac{2\pi i}{N}}{\theta -\frac{2\pi i}{N}}\,\bP_a \,,        
\eal
where \(\bP\) is the permutation operator 
and $\bP_s = {1\ov2}(\bI+\bP)\,,\, \bP_a = {1\ov2}(\bI-\bP)\,,$ are the projection operators 
onto the symmetric and antisymmetric parts of the tensor product of two fundamental representations. The S-matrix $\bS^{\rm PCF}(\t)$ has a single pole at $\theta =\frac{2\pi i}{N}$ due to 
  the pole of $\bR(\theta)$ in the antisymmetric part and the zero of $S_{PCF}(\theta)$ at $\theta =\frac{2\pi i}{N}$ which leads to the existence of the rank-$r$ bound states. 
The $(N$-$1)$-particle bound states are identified with anti-particles of the elementary particles. In general
 a rank-$r$ and a  rank-$(N$-$r)$ particles created by $\cA^\dagger_{K\dK}$ and $\cA^\dagger_{\overline K \overline \dK}$ form a particle-antiparticle pair if ${\overline K}$ and $\overline \dK$ are such that $K \cup{\overline K} =\cP(1,2,\ldots,N)$ and $\dK \cup{\overline \dK} =\dot\cP(\d1,\dt,\ldots,\dN)$ where $\cP$ and $\dot\cP$ are some permutations of  $1,2,\ldots N$, and $\d1,\dt,\ldots,\dN$, respectively. 
In what follows in such a pair a bound state of smaller rank (that is  $r<N/2$) is considered as a particle. If $N$ is even, $N=2 p$, then a bound state with the label $K = (1,k_2,\ldots, k_p)$ is considered as a particle.
The ZF operators can be normalised in such a way that for a particle $\cA^\dagger_{K\dK}$ and antiparticle $\cA^\dagger_{L\dL}$  the charge conjugation matrix $C_{K\dK,L\dL}=\e_{KL}\e_{\dK\dL}$ where $\e_{KL}\equiv\e_{i_1\ldots i_N}$ is skew-symmetric, and $\e_{1\ldots N}=1$, and similarly for $\e_{\dK\dL}$.
The S-matrices of the bound states are obtained from the S-matrix  for elementary particles by the fusion procedure.  The creation and annihilation operators satisfy the ZF algebra \cite{Zam,Fad}.

Up to a CDD factor the \sun2 PCF model can be thought of as the tensor product of two
chiral SU(N) Gross-Neveu models due to the following relation between their S-matrices for the elementary particles
\be
\bS^{\rm PCF}(\theta)= (\bS^{GN}(\theta)\otimes \bS^{GN}(\theta))\chi_{_{CDD}}(-\theta)\, ,
\ee
where
\begin{equation}  \label{SGN-matrix}       
\bS^{\rm GN}(\t) =S_{GN}(\theta)\, {\bR(\theta)}\,,  
\end{equation} 
\bal
S_{GN}(\theta)=S(\theta)\,\chi_{_{CDD}}(\theta)&=\frac{\Gamma \left( \frac{i \theta}{2\pi} \right)       
\Gamma \left(\frac{N-1}{N}- \frac{i \theta}{2\pi} \right)}       
{\Gamma \left(-\frac{i \theta}{2\pi} \right)       
\Gamma \left(\frac{N-1}{N}+ \frac{i \theta}{2\pi} \right)}
\, .
\eal
The \GN S-matrix satisfies the crossing symmetry condition
\bea
\prod
   _{k=-\frac{N-1}{2}}^{\frac{N-1}{2}} S_{GN}(\theta + \frac{2\pi i}{N}\,k) =(-1)^{N-1}\frac{ \theta - i\pi {N-1\ov N}}{\theta + i\pi {N-1\ov N}} \,,
\eea
and has 
the large $\t$ asymptotics
$
S(\pm\infty)=e^{\mp i\pi {N-1\ov N}}\, .
$
It admits the nice integral form
\be
S_{GN}(\theta)=\exp\left(-2\, i\, \int^{\infty}_{0}{dt\ov t} {e^{\pi t\ov N}\,\sinh{(N-1)\pi t \ov N}\ \ov \sinh \pi t}\,\sin\t t\right)\, ,
\ee
Writing an integral representation of the CDD factor, one finds
\be
\chi_{_{CDD}}(\t)=-\exp \left(-2\, i\, \int^{\infty}_{0}{dt\ov t} {\sinh{\pi t\left(1-{2\ov N}\right)} \ov \sinh \pi t}\,\sin\t t\right)\, ,
\ee
and therefore
 \be
S_{PCF}(\theta)=S^{1\dot{1}1\dot{1}}_{1\dot{1}1\dot{1}}(\theta)=\exp\left(-4\, i\, \int^{\infty}_{0}{dt\ov t} {\sinh{\pi t\ov N}\,\sinh\left({(N-1)\pi t \ov N}\right) \ov \sinh \pi t}\,\sin\t t\right) = {g_{PCF}(-\t)\ov g_{PCF}(\t)}\, .
\ee
Here
\be\la{gPCF}
g_{PCF}(\t)=\frac{ \Gamma
   \left(\frac{i \t}{2 \pi }-\frac{1}{N}+1\right) \Gamma
   \left(\frac{i \t}{2 \pi }+\frac{1}{N}\right)}{\Gamma \left(\frac{i
   \t}{2 \pi }\right) \Gamma
   \left(\frac{i \t}{2 \pi }+1\right)}
\ee
is the Green's function which will appear in the free field representation of the PCF model to be constructed in this paper.

%%%%%%%%%%%%%%%%%%%%%%%
\subsection{Form factors}

The ZF operators are used to create the $in$- and $out$-states 
as follows
\bea\nonumber
\hspace{-0.3cm} &&|\t_1,\t_2, \cdots , \t_n
\rangle^{(in)}_{\bI_1,...,\bI_n} =\cA^\dagger_{\bI_n}(\t_n)\cdots  \cA^\dagger_{\bI_1}(\t_1)|vac \rangle
 \,,\quad\qquad  \t_1<\t_2<\cdots <\t_n\,,\\\nonumber
\hspace{-0.3cm} &&|\t_1,\t_2, \cdots , \t_n
\rangle^{(out)}_{\bI_1,...,\bI_n} = \cA^\dagger_{\bI_1}(\t_1)\cdots  \cA^\dagger_{\bI_n}(\t_n)|vac \rangle 
   \,,\quad\qquad  \t_1<\t_2<\cdots <\t_n\,,
\eea 
where $\bI$ is a multi-index $\bI\equiv I\dI$, 
and the vacuum state 
$|vac \rangle$ is annihilated by $\cA^{\bI}(\t)$, and has the unit norm, $\langle vac|vac \rangle=1$.
  
Thus, form factors \eqref{FFdef} of a local operator $O(x)$  can be written as
\be\la{defF}
F_{\bI_1\ldots \bI_n}(\t_1,\dots,\t_n) = \langle vac| O(0) \cA^\dagger_{\bI_n}(\t_n)\cdots  \cA^\dagger_{\bI_1}(\t_1)|vac \rangle\,.
\ee
According to Lukyanov  \cite{Lukyanov}, the determination of form factors can be reduced to the problem of finding a representation of a so-called {\it extended} ZF algebra which is generated by  vertex operators 
$Z_\bI(\t)$, the angular Hamiltonian $\bK$, and the central elements $\Omega_\bI$ 
obeying the defining relations
\bea\la{eZFa}
Z_{\bI}(\t_1)Z_{\bJ}(\t_2)&=& Z_{\bL}(\t_2)Z_{\bK}(\t_1)S_{\bI\bJ}^{\bK\bL}(\t_{12})\,,\\\la{eZFb}
Z_{\bI}(\t_1)Z_{\bJ}(\t_2)&=& -{i\, C_{\bI\bJ}\ov \t_{12}-i\pi}+\cO(1)\,,\quad \t_{12}\to i\pi\,,\\\la{eZFc}
{d\ov d\t}Z_\bI(\t ) &=&-\left[\, \bK\,, \,Z_\bI(\t) \right] 
- i\Omega_\bI Z_\bI(\t)
\,,\\
\la{eZFd}
Z_{\bI}(\t' + i\un_+)Z_{\bJ}(\t-i\un_-)&=& {i\ov \t'-\t}\sum_{\bK\in\cK}\Gamma^\bK_{\bI\bJ}Z_\bK(\t)+\cO(1)\,,\quad \t'\to \t\,,
\eea 
where the notation $\t_{ij}\equiv \t_{i}-\t_{j}$ is used.
The bootstrap conditions \eqref{eZFd} are required if  the particles $\cA^\dagger_\bK$ of the same mass with $\bK\in \cK$ are bound states of particles $\cA^\dagger_\bI$ and $\cA^\dagger_\bJ$ with 
$\bI\in \cI$ and $\bJ\in \cJ$. The mass of the bound state $\cA^\dagger_\bK$ is equal to $m_\bK = m_\bI\cos\un_++m_\bJ\cos\un_-$ where $\un_\pm$ are found from the equations
\be
\un_++\un_-= \un_{\bI\bJ}^\bK\,,\quad m_\bI \sin \un_+ = m_\bJ \sin \un_-\,.
\ee
The scattering matrix of the particles $\cA^\dagger_\bI$ and $\cA^\dagger_\bJ$ must have a simple pole at $\t = i\un_{\bI\bJ}^\bK$.  $\Gamma^\bK_{\bI\bJ}$ are some constants, and
the relations \eqref{eZFd} can be inverted and used to derive the vertex operators for the bound states from the vertex operators for elementary particles.  
 
Notice that due to \eqref{eZFa} and  \eqref{eZFb},  $Z_\bI(\t)$ and 
$C^{\bI\bJ}Z_\bJ(\t+{i\pi})$, $C^{\bI\bJ}C_{\bJ\bK}=\delta^\bI_\bK$ can be thought as representing the ZF creation and annihilation operators, respectively.\footnote{For the SU(2p) chiral GN model the relations  \eqref{eZFb} are modified by replacing $C_{\bI\bJ}$ with $C_{\bI\bJ}\G$ where $\G$ is an auxiliary element satisfying $\G^2=id$ which  (anti-)commutes with $Z_I$ \cite{BF13}.}
 
 An important observation of Lukyanov is that 
 any representation $\pi_O$ of the extended ZF algebra corresponds to a local operator $O$, and the form factors \eqref{defF} of this operator are given by the formula
 \be\la{ffax}
F^O_{\bI_1\ldots \bI_n}(\t_1,\ldots,\t_n)= \cN_O\, {\Tr_{\pi_{O}}\left[e^{2\pi i \,\bK}Z_{\bI_n}(\t_n) \cdots Z_{\bI_1}(\t_1) \right]\ov\Tr_{\pi_{O}}\left[e^{2\pi i \,\bK}\right]}\,,
\ee
where the normalisation constant $ \cN_O$ depends only on the local operator $O$ and has to be fixed by other means. It can be shown that if \eqref{ffax} satisfies the necessary analyticity properties, then the form factor axioms follow from the cyclicity of the trace and the extended ZF algebra.
 
 In addition, if one finds a linear operator $\Lambda(\tilde O)$ acting in $\pi_{O}$ which satisfies 
\be\la{lamO}
e^{\t \,\bK}\Lambda(\tilde O) e^{-\t\, \bK}=e^{\t s(\tilde O)}\Lambda(\tilde O)\,,\quad \Lambda(\tilde O) Z_\bI(\t) = e^{2\pi i\,\Omega(\tilde O,I)}Z_\bI(\t)\Lambda(\tilde O) \,,
\ee
then it corresponds to some  local operator $\tilde O(x)$ with the spin $s(\tilde O)$, and 
the form factors of the operator $:\tilde O(x)O(x):$ are given by  
 \be\la{ffax2}
F^{\tilde OO}_{\bI_1\ldots \bI_n}(\t_1,\ldots,\t_n)= \cN_{\tilde OO}\, {\Tr_{\pi_{O}}\left[e^{2\pi i \,\bK}\Lambda(\tilde O)Z_{\bI_n}(\t_n) \cdots Z_{\bI_1}(\t_1) \right]\ov\Tr_{\pi_{O}}\left[e^{2\pi i \,\bK}\right]}\,.
\ee
Notice that $\Omega(\tilde O,I)$ appears in \eqref{lamO} if the particle $\cA^\dagger_\bI$ has nontrivial statistics with respect
to $\tilde O(x)$.

%%%%%%%%%%%%%%%%%%%%%%%%%%%%%%%%%%%%%%%%%%%%%%%%%%%%
\subsection{Free fields and basic vertex operators}

The second important result by \cite{Lukyanov} is that for many models the extended ZF algebra can be realised in terms of free bosons. 
In what follows free fields
$\p_\mu(\t)$ 
which satisfy the following relations 
\bea\begin{aligned}
\left[\phi_\mu(\t_1), \phi_\nu(\t_2)\right]&=\ln S_{\nu\mu}(\t_{21})\, ,\quad
\langle\phi_\mu(\t_1) \phi_\nu(\t_2)\rangle=-\ln g_{\nu\mu}(\t_{21})\, ,~~~~
\end{aligned}
\eea
are used.
The S-matrices $S_{\mu\nu}$ and Green's functions $g_{\mu\nu}$ are related to  each other as
\be
S_{\mu\nu}(\t)={g_{\nu\mu}(-\t)\ov g_{\mu\nu}(\t)} = {1\ov S_{\nu\mu}(-\t)}\,.
\ee
The fields $\p_\mu$ are used to construct the basic vertex operators
\be
V_\mu(\t)=\, : e^{i\phi_\mu(\t)}:\, \equiv e^{i\phi_\mu(\t)}\,,
\ee
which obey the following  relations
\be
V_\mu(\t_1)V_\nu(\t_2)
=g_{\nu\mu}(\t_{21}):V_\mu(\t_1)V_\nu(\t_2):\, ,\quad
V_\mu(\t_1)V_\nu(\t_2)
=S_{\mu\nu}(\t_{12})V_\nu(\t_2)V_\mu(\t_1)\, .\la{VVrel}
\ee
The free fields can be written in the form
\bal\label{FieldInt}
\phi_\m(\theta)
&=Q_\mu+\int^\infty_{-\infty}{dt\over i t}\,
A_\m(t)e^{i\theta t}\, ,
\eal
where the creation $A_\m(-t)\,,\ t>0$ and annihilation operators $A_\m(t)|0\ket =0\,,\ t>0$ have the commutation relations 
\be\la{comrel}
[A_\mu(t),A_\nu(t')]=t f_{\nu\mu}(t)\delta(t+t')\,,\quad f_{\mu\nu}(-t)=f_{\nu\mu}(t)\,.
\ee
The zero mode operators $Q_\mu$ commute with $A_\nu(t)$, and are introduced to guarantee that the ZF operators have correct eigenvalues with respect to the Cartan generators $P_\mu$ which annihilate the vacuum: $P_\mu |0\rangle = 0$. Their  commutation relations 
will not be important for this paper.

The S-matrices $S_{\mu\nu}$ and Green's functions $g_{\mu\nu}$ are related to $f_{\mu\nu}$ as follows
\be
S_{\mu\nu}(\t)=\exp\left(\int_{-\infty}^\infty\, {dt\ov t}\, f_{\mu\nu}(t)e^{-i\theta t}\right)\,, \quad g_{\mu\nu}(\t)=\exp\left(-\int_{0}^\infty\, {dt\ov t}\, f_{\mu\nu}(t)e^{-i\theta t}\right)\,,
\ee
where it is assumed that the integral representation for $S_{\m\nu}(\t)$ is well-defined for some $\t$'s. If it is not defined then one should analytically continue Green's functions $g_{\mu\nu}(\t)$ and $g_{\nu\mu}(-\t)$ to a common domain and calculate their ratio to find $S_{\m\nu}(\t)$.

The annihilation and creation parts of $\p_\mu(\t)$ will be denoted as follows
\bal\label{Fieldpm}
\phi_\m^-(\theta)
=\int_0^{\infty}{dt\over i t}\,
A_\m(t)e^{i\theta t}\, ,
\quad
\phi_\m^+(\theta)
&=\int^0_{-\infty}{dt\over i t}\,
A_\m(t)e^{i\theta t}=-\int_0^{\infty}{dt\over i t}\,
A_\m(-t)e^{-i\theta t}\,.
\eal

\begin{figure}[t]
\begin{center}
\includegraphics[width=0.36\textwidth]{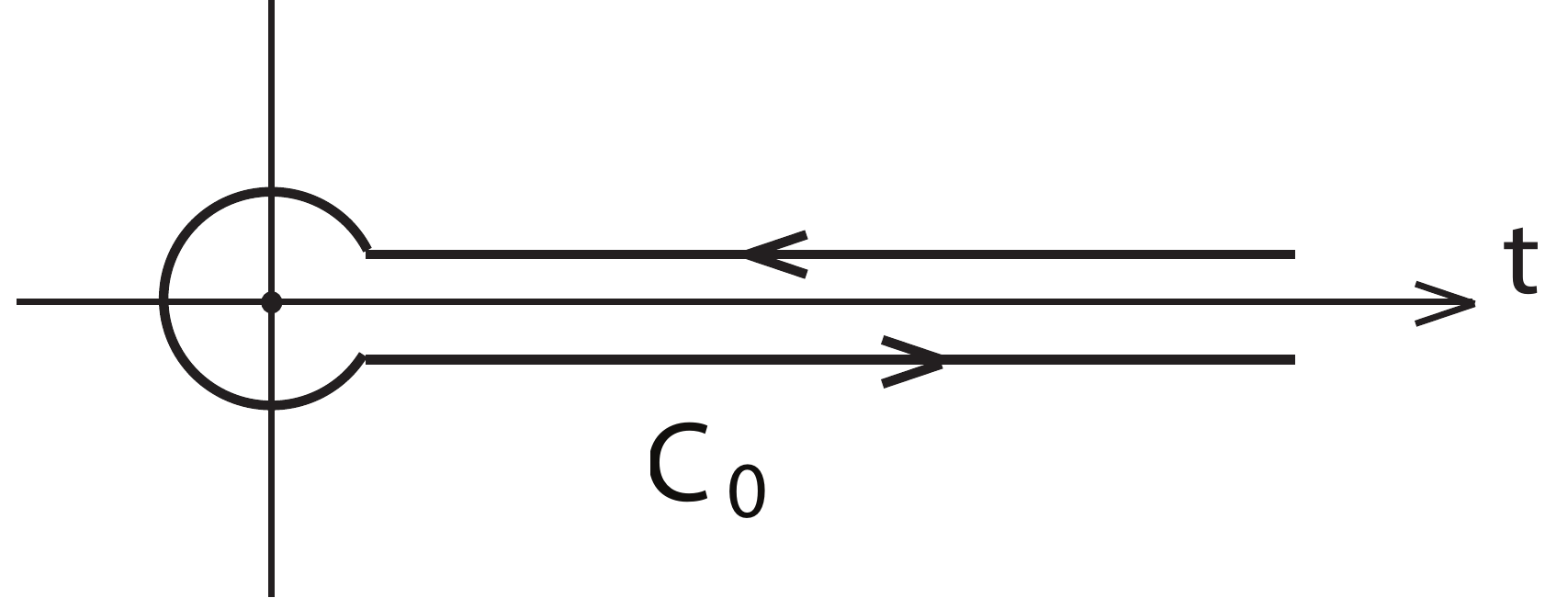}
\caption{\small The integration contour $C_0$ in the integral $\int_{C_0}\, {dt\ov 2\pi i}\, F(t)\, \ln(-t)$.}
\label{C0}
\end{center}
\end{figure}

It is also important to mention that  the integrals of the form 
\be
\int_{0}^\infty\, {dt}\, F(t)
\ee
will be always understood as \cite{JKM}
\be
\int_{0}^\infty\, {dt}\, F(t) \equiv 
\int_{C_0}\, {dt\ov 2\pi i}\, F(t)\, \ln(-t)\,,
\ee
where the integration contour $C_0$ is shown in Figure \ref{C0}. 

%%%%%%%%%%%%%%%%%%%%%%%%%%%%%%%%
\section{Free field representation} \la{free}

The ZF operators for the elementary particles of the PCF model are $\bA_{k\dk}^\dagger$, $k=1,\ldots, N$, $\dk = \d1 , \ldots,\dN$.\footnote{The dotted and undotted indices are obviously different $\dk \neq k$. However, in some formulae they are identified which will be clearly stressed out.}
The \sun2 charges $\bR_{b}{}^a$ and $\bR_{\db}{}^{\da}$ act on the ZF operators as follows
\bal\la{JAs2b}
 \bR_b{}^a\, \bA_{k\dk}^\dagger -\bA_{k\dk}^\dagger\,
\bR_b{}^a &=\de_k^a\,\bA_{b\dk}^\dagger\,,\quad
\dot\bR_{\db}{}^{\dot a}\, \bA_{k\dk}^\dagger -\bA_{k\dk}^\dagger\,
\dot\bR_{\db}{}^{\dot a} =\de_\dk^\da\bA_{k\db}^\dagger\,.
\eal
It is natural to use $\bA_{1\d1}^\dagger$ as the highest weight operator, and charges $\bJ_k\equiv \bR_{k+1}{}^k$ and $\dot\bJ_\dk\equiv \dot\bR_{\dk +\d1}{}^{\dk}$ as the lowering operators 
which  produce all elementary particles ZF operators from $\bA_{1\d1}^\dagger$.

%%%%%%%%%%%%%%%%
\subsection{Charges and free fields in the large $N$ limit}  \la{free1}

To motivate the free field representation which is described in the next subsection it is useful to consider the large $N$ limit. In this limit the S-matrix goes to the identity matrix, and the ZF algebra for the ZF operators becomes the usual Heisenberg algebra of the annihilation and creation operators $a^{m\dm}$ and $a_{n\dn}^\dagger$ whose nontrivial commutation relations are
\bal
a^{m\dm}(\t_1)a_{n\dn}^\dagger(\t_2) -a_{n\dn}^\dagger(\t_2) a^{m\dm}(\t_1) = \de^m_n\de^\dm_\dn\de(\t_{12})\,.
\eal
These operators and relations depend on the rapidities  but the dependence will be often omitted to simplify the expressions below. The charges of the two copies of $\su(N)$'s are built from the operators
\bal
V_{m\dr}^n\equiv a_{m\dr}^\dagger a^{n\dr} \,,
\quad \dV_{\dm r}^{\dn}\equiv a_{r\dm}^\dagger a^{r\dn}\,,
\eal
These operators satisfy the commutation relations
\bal
V_{m\dr}^n V_{k\dq}^l - V_{k\dq}^l V_{m\dr}^n=\de_k^n\de_{\dq\dr} V_{m\dr}^l +\de_m^l\de_{\dq\dr} V_{k\dr}^n\,,
\eal
\bal
\dV_{\dm r}^{\dn}\dV_{\dk q}^{\dl}-\dV_{\dk q}^{\dl}\dV_{\dm r}^{\dn}=\de_\dk^\dn\de_{qr} \dV_{\dm r}^{\dl} +\de_\dm^\dl\de_{qr} \dV_{\dk r}^{\dn}\,,
\eal
\bal
V_{m\dr}^n \dV_{\dk q}^{\dl}- \dV_{\dk q}^{\dl}V_{m\dr}^n=(\de_{q}^{n}\de_\dk^\dr -\de_{m}^{q}\de_\dr^\dl )a_{m\dk}^\dagger a^{n\dl}\,.
\eal
One sees from the last relation that the operators
\bal
R_m^n\equiv \sum_{\dr=\d1}^\dN \int d\a\,V_{m\dr}^n(\a) \,,\qquad \dR_{\dk}^{\dl}\equiv \sum_{q=1}^N \int d\a\,\dV_{\dk q}^{\dl}(\a) \,,
\eal
commute
\bal
R_{m}^n \dR_{\dk}^{\dl}-\dR_{\dk}^{\dl}R_{m}^n=0\,,
\eal
and it is not difficult to check that they form the $\un(N)\oplus\un(N)$ algebra.

The charges  $R_{k+1}{}^k$ and 
$R_{\dk+\d1}{}^{\dk}$ are  linear combinations
of
$
 V_{k+1,\dr}^k$ and  $\dV_{\dk+\d1,r}^{\dk}$ 
whose nontrivial  relations are
\bal\la{VkVn}
& V_{k+1,\dr}^k V_{n\dq}^{n-1}=\de_{kn}\de_{\dr\dq} V_{k+1,\dr}^{k-1}+:V_{k+1,\dr}^k V_{n\dr}^{n-1}:\,,
\eal
\bal\la{VdkVdn}
& \dV_{\dk+\d1,r}^\dk \dV_{\dn q}^{\dn-\d1}=\de_{\dk\dn}\de_{rq} \dV_{\dk+\d1,r}^{\dk-\d1}+:\dV_{\dk+\d1,r}^\dk \dV_{\dn q}^{\dn-\d1}:\,,
\eal
\bal\la{VkVdn}
&  V_{k+1,\dr}^k \dV_{\dn q}^{\dn-\d1}=\de_{\dr\dn}\de_{kq} a_{k+1,\dn}^\dagger a^{k,\dn-\d1}+: V_{k+1,\dr}^k \dV_{\dn q}^{\dn-\d1}:\,,
\eal
\bal\la{VdkVn}
&  \dV_{\dn q}^{\dn-\d1}V_{k+1,\dr}^k =\de_{\dr,\dn-\d1}\de_{k+1,q} a_{k+1,\dn}^\dagger a^{k,\dn-\d1}+: \dV_{\dn q}^{\dn-\d1}V_{k+1,\dr}^k:\,,
\eal
where 
Kronecker's deltas are multiplied by Dirac's delta of the difference of the rapidities the operators depend on.

In terms of free fields it is therefore reasonable to assume that the charges $\bJ_k$ and $\dot\bJ_\dk$ are linear combinations
of the following vertex operators
\bal
 &V_{k+1,\dr}^k(\a)\sim e^{i\p_{k\dr}(\a)}\,,\quad \dV_{\dk+\d1,r}^{~\dk}(\a)\sim e^{i\dphi_{\dk r}(\a)}\,.
\eal
Moreover, it is natural to expect that a delta-function term in the product $V_\mu(\t_1)V_\nu(\t_2)$ of two $V$'s
in (\ref{VkVn}-\ref{VdkVn}) implies that the corresponding Green's function $g_{\nu\mu}(\t_{21})$ has a pole at $\t_{21}=0$. On the other hand even if the product of two $V$'s in (\ref{VkVn}-\ref{VdkVn}) is regular it does not mean that the corresponding Green's function is equal to 1 because these relations are only valid in the large $N$ limit. 

Thus, 
the appearance of the delta-function terms in the products
$$V_{k+1,\dr}^k V_{k\dr}^{k-1}\,,\quad \dV_{\dk+\d1,r}^\dk \dV_{\dk r}^{\dk-\d1}\,,\quad V_{k+1,\dn+\d1}^k \dV_{\dn+\d1,k}^{\dn}\ {\rm\ and\ }\ \dV_{\dn+\d1,k+1}^{\dn}V_{k+1,\dn}^k$$ implies that the  Green's functions 
$$g_{k-1,\dr|k\dr}(\t)\,,\quad g_{\dk-\d1,r|\dk r}(\t)\,,\quad g_{\dn k|k,\dn+\d1}(\t)\ {\rm\ and\ }\ g_{k\dn|\dn,k+1}(\t)$$
have a simple pole at $\t=0$. 

Then, the commutativity of the left and right charges suggests that
not all of the $2(N-1)N$ free fields are independent, and one should have the following relations
\bal
\p_{k,\dn}+\dphi_{\dn,k+1}=\p_{k,\dn+\d1}+\dphi_{\dn,k}\,,\quad k=1,\ldots,N-1\,,\  \dn=\d1,\ldots,\dN-\d1\,.
\eal 
Thus, it is expected that the number of independent fields is $N^2-1$ which matches the number of fields in the $SU(N)$ PCF Lagrangian.

Notice that the zero modes $Q_{k\dr}$, $Q_{\dk r}$, $P_{k\dr}$ and  $P_{\dk r}$ should  satisfy
\bal
Q_{k\dr} = Q_k\,,\quad  P_{k\dr} = P_k\,,\quad Q_{\dk r}=Q_\dk\,,\quad  P_{\dk r}=P_\dk, 
\eal
where the zero modes $P_k$, $Q_k$ and $P_\dk$, $Q_\dk$ commute with the oscillators in the fields, and  satisfy the algebra
\be
\left[P_k, Q_n\right]=i a_{kn}\,, \quad \left[Q_k, Q_n\right]=\left[P_k, P_n\right]=0\,,\quad k,n=0,1,\ldots,N\,.
\ee
Here $a_{ij}=2 \delta_{ij} - \delta_{i-1,j} - \delta_{i+1,j}$ is the Cartan matrix of type $A_{N-1}$.
Similar relations hold between $P_\dk$, $Q_\dk$. In addition the dotted and undotted zero modes commute.

%%%%%%%%%%%%%%%%
\subsection{Ansatz for ZF operators and charges}

The ZF operators $\bA_{k\dr}^\dagger$ and the lowering operators $\bJ_k^-$, $\dot\bJ_{\dk}^-$
will be represented by the ZF vertex operators $Z_{k\dr}$ and charges $\chi_k^-$, $\dot\chi_{\dk}^-$.
According to the discussion above, a rather general ansatz is
\bal\la{ZFop}
Z_{1\d1}(\t)&=\r(\t) e^{i\p_0(\t)}\,,\\
  \chi_k^-&=\sum_{\dr=\d1}^\dN\int_{C_{k\dr} }{d\a_{k\dr}\ov 2\pi} \,c_{k\dr}(\a_{k\dr})e^{i\p_{k\dr}(\a_{k\dr})}\,,\quad \dot\chi_\dk^-=\sum_{r=1}^N\int_{C_{\dk r}}{d\a_{\dk r}\ov 2\pi} \,c_{\dk r}(\a_{\dk r})e^{i\dphi_{\dk r}(\a_{\dk r})}\,,\\
Z_{k+1,\dr}(\t)&=\lok \,Z_{k\dr}(\t)-Z_{k\dr}(\t)\,\lok  \,,\quad k=1,\ldots,N-1\,,
\\
Z_{k,\dr+1}(\t)&=\dot\chi_{\dr}^- \,Z_{k\dr}(\t)-Z_{k\dr}(\t)\,\dot\chi_{\dr}^-  \,,\quad \dr=\d1,\ldots,\dN-\d1\,,
\eal
where the functions $\r(\t)$, $c_{k\dr}(\a)$ and $c_{\dk r}(\a)$ satisfy some relations to be determined in the following sections.
The free field $\p_0$ can be expressed in terms of $\p_{k\dr}\,,\,\dphi_{\dk r}$ due to the relations \eqref{eZFc} of the extended ZF algebra, see \eqref{popkr}.
The integration contour $C$ in any operator $\chi$ which involves integration is fixed as follows  \cite{Lukyanov}.  The product of all vertex operators in a monomial containing $\chi$ is normal ordered. This produces a product of various Green's functions. The contour $C$ runs from Re$\,\a=-\infty$ to Re$\,\a=+\infty$ and it lies above all poles due to operators to the right of $\chi$ but below all poles due to operators to the left of $\chi$. 
The contour $C$ should be additionally deformed according to the procedure described if one then acts by the resulting monomial operator on other operators.

Recall, that 
\be
V_\mu(\t_1)V_\nu(\t_2)
=g_{\nu\mu}(\t_{2}-\t_1):V_\mu(\t_1)V_\nu(\t_2):\,,
\ee
and $\mu\,,\ \nu$ can be $0$, or $k\dr$ or $\dk r$. Thus, one has the following Green's functions
\bal
g_{00}=g_{PCF}\,,\quad g_{0|k\dr}\,,\ g_{k\dr|0}\,,\quad  g_{0|\dk r}\,,\ g_{\dk r|0}\,,\quad
 g_{k \dr|n \dq}\,,\quad g_{\dk r|\dn q}\,,\quad
 g_{k \dr|\dn q}\,,\ g_{\dn q|k \dr}\,, 
\eal 
and the corresponding S-matrices. The Green's functions are listed in subsection \ref{listGreen}, and their derivation is sketched in appendix \ref{Greender}.
In what follows $\p_{k\dr}$ and $\p_{\dk r}$ are sometimes referred to as the left and right sector fields, respectively. Their Green's functions $g_{k \dr|n \dq}\,,\, g_{\dk r|\dn q}\,,
\, g_{k \dr|\dn q}\,,\, g_{\dn q|k \dr}$ are referred to as the left-left, right-right, left-right, and right-left functions. 
 
By using these formulae, one finds in particular
\bal
Z_{2\d1}(\t)&=\chi_1^-\,Z_{1\d1}(\t)-Z_{1\d1}(\t)\,\chi_1^- = \r(\t)\sum_\dr\int_{C_{1\dr}}{d\a_{1\dr}\ov 2\pi} \,c_{1\dr}(\a_{1\dr})g^a_{0|1\dr}(\t-\a_{1\dr})e^{i\p_0(\t)+i\p_{1\dr}(\a_{1\dr})}\,,
\eal
\bal
Z_{1\dt}(\t)&= \dot\chi_\d1^-\,Z_{1\d1}(\t)-Z_{1\d1}(\t)\, \dot\chi_\d1^- = \r(\t)\sum_r\int_{C_{\d1 r}}{d\a_{\d1 r}\ov 2\pi} \,c_{\d1 r}(\a_{\d1 r})g^a_{0|\d1 r}(\t-\a_{\d1 r})e^{i\p_0(\t)+i\dot\p_{\d1 r}(\a_{\d1 r})}\,,
\eal
where the integration contours run above the poles of $g_{0|A}$ functions, and below the poles of $g_{A|0}$ functions, and  for any Green's function one defines
\bal
g^a_{A|B}(\a-\b) \equiv g_{A|B}(\a-\b)-g_{B|A}(\b-\a)\,.
\eal
To simplify formulae the summation symbols and the differentials ${d\a_{k\dr}\ov 2\pi}$,  ${d\a_{\dk r}\ov 2\pi}$, and the dependance of free fields and Green's functions on their arguments are often dropped: $\p_A\equiv \p_A(\a_A)$,  $g_{A|B}\equiv g_{A|B}(\a_A-\a_B)$, e.g.
\bal
&\p_0\equiv \p_0(\t)\,,\quad \p_{k\dr}\equiv \p_{k\dr}(\a_{k\dr})\,,\quad \dot\p_{\dk r}\equiv \dot\p_{\dk r}(\a_{\dk r})\\
&g_{0|1\dr}\equiv g_{0|1\dr}(\t-\a_\dr)\,,\quad g_{\dk q|n\dr}\equiv g_{\dk q|n\dr}(\a_{\dk q}-\a_{n\dr}) \,,
\eal
unless there is an ambiguity. 

 The ZF operator $Z_{m\dn}$ then can be symbolically written as follows
\bal\la{Zmdn}
Z_{m\dn}(\t)&=\r(\t)\int \prod_{k=1}^{m-1} c_{k\dr_k}g^a_{k-1,\dr_{k-1}|k \dr_k}\prod_{\dk=\d1}^{\dn-\d1} c_{\dk r_\dk}g^a_{\dk-\d1,r_{\dk-\d1}|\dk r_\dk}
\prod_{k=1}^{m-1}\prod_{\dk=\d1}^{\dn-\d1}g^a_{\dk r_{\dk}|k \dr_k}\\
&\times \exp({i\p_0+i\sum_{k=1}^{m-1}\p_{k\dr_k}+i\sum_{\dk=\d1}^{\dn-\d1}\dot\p_{\dk r_\dk}})\,,
\eal
where $g^a_{0\dr_{0}|1\dr_1}\equiv g^a_{0|1\dr_1}$, the sum over $\dr_k$ and $r_\dk$ is taken, and in each term of the sum the integration contours run according to the rule decribed above.

%%%%%%%%%%
\subsection{ List of Green's functions } \la{listGreen}

Here are all Green's functions different from 1  collected.
\bal
 g_{0|1\d1}(\t)&= g_{0|\d1 1}(\t)={ \t-{2\pi i\ov N}\ov\t}\,, \quad g_{1\d1|0}(\t)=g_{\d1 1|0}(\t)=1\,, 
\\
g_{0|1\dr}(\t)&=g_{0|\d1 r}(\t)=1\,,\quad g_{1\dr|0}(\t)=g_{\d1 r|0}(\t)=\frac{\theta }{\theta+\frac{2\pi i}{N} }\,,\quad  
\d1<\dr\,,\ 1<r\,,
\eal
\be
g_{k\dr|k\dq}(\t)=g_{\dk r|\dk q}(\t)=1\,,\quad g_{k\dq|k\dr}(\t)=g_{\dk q|\dk r}(\t)={\theta+\frac{2\pi i}{N}\ov \theta -\frac{2\pi i}{N}}\,,\quad \dr< \dq\,,\ r< q\,,
\ee
\be
g_{k\dr|k\dr}(\t)=g_{\dk r|\dk r}(\t)={\theta\ov \theta -\frac{2\pi i}{N}}\,,\quad \dk\,,\, \dr\ge\d1\,,\ k\,,\, r\ge 1\,,
\ee
\bal
g_{k\dr|k+1,\dq}(\t)&=g_{\dk r|\dk+\d1,q}(\t)={ \t-{2\pi i\ov N}\ov\t}\,, \quad g_{k+1,\dq|k\dr}(\t)=g_{\dk+\d1,q|\dk r}(\t)=1\,, \quad\dr\ge \dq\,,\ r\ge q\,,
\\
g_{k\dr|k+1,\dq}(\t)&=g_{\dk r|\dk+\d1,q}(\t)=1\,, \quad g_{k+1,\dq|k,\dr}(\t)=g_{\dk+\d1,q|\dk r}(\t)=\frac{\t}{\t+\frac{ 2\pi i}{N} }\,, \quad\dr< \dq\,,\ r< q\,,
\eal
\be
g_{k\dr|\dr, k+1}(\t)=g_{\dr k|k, \dr+\d1}(\t)={\theta-\frac{2\pi i}{N}\ov \theta }\,,\quad 
g_{\dr, k+1|k\dr}(\t)=g_{k, \dr+\d1|\dr k}(\t)={\theta+\frac{2\pi i}{N}\ov \theta }\,.
\ee
The Green's functions satisfy the following equations
\bal\la{Greenfrel}
&g_{A|k\dr}(\t)g_{A|\dr,k+1}(\t)=g_{A|\dr k}(\t)g_{A|k,\dr+\d1}(\t) \,,\quad g_{k\dr|A}(\t)g_{\dr,k+1|A}(\t)=g_{\dr k|A}(\t)g_{k,\dr+\d1|A}(\t)\,,
\eal
where $A$ is any of the indices of the fields. These equations are obtained 
from the following constraints between the free fields
\bal\la{fieldrel1}
\p_{k\dr}(\t)+\dphi_{\dr,k+1}(\t)=\dphi_{\dr k}(\t)+\p_{k,\dr+\d1}(\t)\,.
\eal 
Then, the functions $c_{k\dr}(\a)$ and $c_{\dk r}(\a)$ satisfy the equations
\bal\la{fieldrel2}
&c_{k\dr}(\t)c_{\dr, k+1}(\t)=c_{\dr k}(\t)c_{k,\dr+\d1}(\t)\,.
\eal
Both \eqref{fieldrel1} and \eqref{fieldrel2} are derived from the commutativity of the left and right algebras charges.

The constraints \eqref{fieldrel1} are solved in the appendix \ref{solvconstr}, and the fields $\p_{k\dr}\,,\, \dphi_{\dr k}$ and $\p_0$ are expressed in terms of the $N^2-1$ elementary free fields defined in \eqref{elemff}.

\medskip

It is worthwhile to mention that shifting the variables $\a_{k\dr}$ as
\bal
\a_{k\dr}\to \a_{k\dr} -{\pi i\ov N} k\,,\quad \a_{\dk r}\to \a_{\dk r} -{\pi i\ov N} \dk\,,
\eal
one can transform all left-left and right-right functions to the ones which have zeroes and poles at the same locations as the \GN functions. However, the left-right Green's functions would have zeroes and poles at positions which depend on their indices. To be precise
\be
g_{k\dr|\dr, k+1}(\t)\to{\theta-\frac{\pi i}{N}(2+k-\dr)\ov \theta-\frac{\pi i}{N}(k-\dr) }\,,\quad 
g_{\dr, k+1|k\dr}(\t)\to{\theta+\frac{\pi i}{N}(2+k-\dr)\ov \theta+\frac{\pi i}{N}(k-\dr) }\,,
\ee
\be
g_{\dr k|k, \dr+\d1}(\t)\to{\theta-\frac{\pi i}{N}(2+\dr-k)\ov \theta-\frac{\pi i}{N}(\dr-k) }\,,\quad 
g_{k, \dr+\d1|\dr k}(\t)\to{\theta+\frac{\pi i}{N}(2+\dr-k)\ov \theta+\frac{\pi i}{N}(\dr-k) }\,.
\ee

%%%%%%%%%%%%%%%%%%%%%%%%%%%
\subsection{The angular Hamiltonian}
An important step in constructing a free field representation is to find an angular Hamiltonian.
The most general Hamiltonian is of the form
\be
\bK=i\,\int^{\infty}_{0}dt \, \sum_{A, B} h_{BA}(t) a_{A}(-t) a_{B}(t)\, ,
\ee
where the sum runs over all independent elementary operators.
The relations to be satisfied are 
\be\la{eZFc2}
{d\ov d \t} Z_{m\dn}(\t)=-\left[\,\bK,Z_{m\dn}(\t)\right]- i\Omega_{m\dn}Z_{m\dn}(\t)\, .
\ee
Computing the derivative of \eqref{Zmdn} with respect to $\t$ it is straightforward to show that if
\bal\la{rhocc}
\r(\t)=e^{i\varkappa\,\t}\tilde\r\,,\quad c_{k\dr}(\t)=e^{i\vk_{k} \t}\tilde c_{k\dr}\,,\quad c_{\dk r}(\t)=e^{i\vk_{\dk} \t}\tilde c_{\dk r}\,,\quad
\eal
 then
\bal
{d\ov d \t}Z_{m\dn}(\t)&=-i\Omega_{m\dn}Z_{m\dn}(\t)
\\
&+
\r(\t)\int \prod_{k=1}^{m-1} c_{k\dr_k}g^a_{k-1,\dr_{k-1}|k \dr_k}\prod_{\dk=\d1}^{\dn-\d1} c_{\dk r_\dk}g^a_{\dk-\d1,r_{\dk-\d1}|\dk r_\dk}
\prod_{k=1}^{m-1}\prod_{\dk=\d1}^{\dn-\d1}g^a_{\dk r_{\dk}|k \dr_k}\\
&\times \big({\pa\ov \pa\t}+\sum_{k=1}^{m-1}{\pa\ov \pa\a_{k\dr_k}}+\sum_{\dk=\d1}^{\dn-\d1}{\pa\ov \pa\a_{\dk r_k}}\big) \exp\big({i\p_0+i\sum_{k=1}^{m-1}\p_{k\dr_k}+i\sum_{\dk=\d1}^{\dn-\d1}\p_{\dk r_\dk}}\big)\,,~~~~~~~
\la{dZmdn}
\eal
where $\vk\,,\,\vk_k\,,\,\vk_\dk$, $\tilde\r\,,\,\tilde c_{k\dr}\,,\,\tilde c_{\dk r}$ are constants, and
\bal\la{Ommdn}
\Omega_{m\dn} = -\big({\vk+\sum_{k=1}^{m-1}\vk_{k}+\sum_{\dk=\d1}^{\dn-\d1}\vk_{\dk}}\big)\,.
\eal
Thus it is sufficient to find $\bK$ such that 
\be
\left[\,\bK,V_\mu(\t)\right]=- {d\ov d \t} V_\mu(\t)\, ,
\ee
where $V_\mu$  is any vertex operator
\be
V_\mu(\t)=e^{i\p_\mu(\t)}\,,\quad \p_\mu(\t)=\int_{-\infty}^\infty\, {dt\ov it}\,\Phi_{\mu A}(t)\,a_A(t)\, e^{i\t t}
\ee
where the index $A$ runs over the indices of the $N^2-1$ independent elementary operators.
One gets
\be
-{d\ov d \t} V_\mu(\t)=-i\int_{-\infty}^\infty\, dt\, e^{i\t t}:\Phi_{\mu A}(t)\,a_A(t)\, e^{i\p_\mu(\t)}:\,,
\ee
and
\bal
\left[\,\bK,V_\mu(\t)\right]=:[\,\bK,i\p_\mu(\t)]e^{i\p_\mu(\t)}:&= -i\,\int_{-\infty}^0\, dt\, e^{i\t t}\Phi_{\mu B}(t)\, h_{BA}(-t):a_A(t)e^{i\p_\mu(\t)}:\\
& -i\, \int_{0}^\infty\, dt\, e^{i\t t}\Phi_{\mu B}(t)\,h_{AB}(t):a_A(t)e^{i\p_\mu(\t)}: \,.
\eal
where it is used that $f_{AB}(t)=\de_{AB}$.
Thus, $h_{AB}(t)=\de_{AB}$, and the angular Hamiltonian is given by the following simple formula
\be
\bK=i\,\int^{\infty}_{0}dt \, \sum_{A} a_{A}(-t) a_{A}(t)\,.
\ee

%%%%%%%%%%%%%%%%%%%%%%%%%%%%%%%%%%%%%%%%%%%
\subsection{Commutativity relations} \la{commutrel}

The charges and ZF operators must satisfy the commutativity relations
\bal\la{chichi}
[\chi^-_k\,,\, \chi^-_n]=0  \,,\quad |k-n|\neq 1\,,
\eal
\bal\la{chichichi}
[[\chi^-_k\,,\, \chi^-_{k+1}]\,,\, \chi^-_{k}]=0 \,,\quad [[\chi^-_k\,,\, \chi^-_{k+1}]\,,\, \chi^-_{k+1}]=0 \,,
\eal
\bal\la{dchidchi}
[\dot\chi^-_\dk\,,\, \dot\chi^-_\dn]=0\,,\quad |\dk-\dn|\neq \d1\,,
\eal
\bal\la{dchidchidchi}
[[\dot\chi^-_\dk\,,\, \dot\chi^-_{\dk+\d1}]\,,\, \dot\chi^-_{\dk}]=0\,,\quad [[\dot\chi^-_\dk\,,\, \dot\chi^-_{\dk+\d1}]\,,\, \dot\chi^-_{\dk+\d1}]=0\,,
\eal
\bal\la{chidchi}
[\chi^-_k \,,\,\dot\chi^-_\dn]=0 \,, \quad \forall\ k\,,  \dn\,,
\eal
\bal\la{chiZZchi}
\chi^-_k \,Z_{m\dn}(\t)=Z_{m\dn}(\t)\,\chi^-_k  \,,\quad \d1\le \dn \le\dN\,,\quad  {\rm unless}\quad m=k\,,
\eal
\bal\la{dchiZZdchi}
\dot\chi^-_\dk \,Z_{n\dm}(\t)=Z_{n\dm}(\t)\,\dot\chi^-_\dk\,,\quad \quad  1\le n\le N\,,\quad  {\rm unless}\quad \dm=\dk\,,
\eal
which follow from the $\su(N)\oplus\su(N)$ symmetry algebra of the model.

The relations \eqref{chichi} and \eqref{dchidchi} just imply that the Green's functions $g_{k\dr|l\dq}(\t)=g_{\dk r|\dl q}(-\t)$ have no poles unless $|k-l|=1$, $|\dk-\dl|=\d1$. The simplest choice used in this paper is $g_{k\dr|l\dq}=g_{\dk r|\dl q}=1$. Then, the commutativity \eqref{chidchi} of the left and right charges  leads to the relations \eqref{fieldrel1} and \eqref{fieldrel2}, and together with  \eqref{chichi} and \eqref{dchidchi} guarantees that \eqref{chiZZchi} and \eqref{dchiZZdchi} hold as soon as the relations 
\bal\la{chiZZchi2}
\chi^-_k \,Z_{k+1,\d1}(\t)&=Z_{k+1,\d1}(\t)\,\chi^-_k  \,,\quad \chi^-_k \,Z_{k+2,\d1}(\t)=Z_{k+2,\d1}(\t)\,\chi^-_k  \,,\\
 \dot\chi^-_\dk \,Z_{1,\dk+\d1}(\t)&=Z_{1,\dk+\d1}(\t)\,\dot\chi^-_\dk\,,\quad \dot\chi^-_\dk \,Z_{1,\dk+\dt}(\t)=Z_{1,\dk+\dt}(\t)\,\dot\chi^-_\dk\,,
\eal
are satisfied. The relations \eqref{chiZZchi2} for $k>1$ follow from \eqref{chichichi} and \eqref{dchidchidchi} which can be easily verified. The commutativity of $\chi_1^-$ with $Z_{2\d1}$, and $\dchi_\d1^-$ with $Z_{1\dt}$ is proven in next section.

%%%%%%%%%%%%%%%%
\section{The ZF algebra} \la{zfalg}

\subsection{The ZF relations}

The ZF algebra relations for $Z_{k\dr}$, $Z_{l\dq}$ take the form
\bal\la{zzkdrldq}
Z_{k\dr}(\t_1) Z_{l\dq}(\t_2) =S(\t_{12})\Big[&s_{12}^2 Z_{k\dr}(\t_2) Z_{l\dq}(\t_1)+s_{21}^2 Z_{l\dq}(\t_2)Z_{k\dr}(\t_1) \\
&+s_{12}s_{21}\big(Z_{k\dq}(\t_2) Z_{l\dr}(\t_1)+Z_{l\dr}(\t_2) Z_{k\dq}(\t_1)\big)\Big]\,.
\eal
where 
\bal
s_{12}= -{{2\pi i\ov N}\ov \t_{12}-{2\pi i\ov N}}\,,\quad s_{21}= {\t_{12}\ov \t_{12}-{2\pi i\ov N}} \,,\quad s_{12}+s_{21}=1\,.
\eal
The relations simplify  for $\dq=\dr$
\be\la{zzkdrldr}
Z_{k\dr}(\t_1) Z_{l\dr}(\t_2) =S(\t_{12})\left[s_{12} Z_{k\dr}(\t_2) Z_{l\dr}(\t_1)+s_{21} Z_{l\dr}(\t_2) Z_{k\dr}(\t_1)\right]\,,
\ee
and  for $l=k$
\be\la{zzkdrkdq}
Z_{k\dr}(\t_1) Z_{k\dq}(\t_2) =S(\t_{12})\left[s_{12} Z_{k\dr}(\t_2) Z_{k\dq}(\t_1)+s_{21} Z_{k\dq}(\t_2) Z_{k\dr}(\t_1)\right]\,,
\ee
and they take the simplest form for $l=k$, $\dq=\dr$
\be\la{zzkdrkdr}
Z_{k\dr}(\t_1) Z_{k\dr}(\t_2) =S(\t_{12})Z_{k\dr}(\t_2) Z_{k\dr}(\t_1)\,.
\ee
Up to the S-matrix $S(\t_{12})$, the relations \eqref{zzkdrldr} or \eqref{zzkdrkdq} are the same as the ZF algebra relations of the \GN model.

All these ZF relations follow from the commutativity relations discussed in subsection \ref{commutrel}, and the ZF algebra relations  for $Z_{1\d1}$, $Z_{2\d1}$, $Z_{1\dt}$, $Z_{2\dt}$. Indeed,
assume that the ZF relations \eqref{zzkdrldq} have been proven for all indices which are less or equal to  $k$, $l$, $\dr$, $\dq$. It is then easy to prove that the ZF relations hold for $k+1$, $l$, $\dr$, $\dq$. Indeed, for $l\neq k$ one gets
\bal\la{zzk1drldq}
Z_{k+1,\dr}(\t_1) Z_{l\dq}(\t_2) &= (\chi_k^-Z_{k\dr}(\t_1) -Z_{k\dr}(\t_1)\chi_k^- ) Z_{l\dq}(\t_2)\\
&= \chi_k^-Z_{k\dr}(\t_1)Z_{l\dq}(\t_2)  -Z_{k\dr}(\t_1)Z_{l\dq}(\t_2) \chi_k^-  
\\
&=S(\t_{12})\Big[s_{12}^2 Z_{k+1,\dr}(\t_2) Z_{l\dq}(\t_1)+s_{21}^2 Z_{l\dq}(\t_2)Z_{k+1,\dr}(\t_1) \\
&\qquad+s_{12}s_{21}\big(Z_{k+1,\dq}(\t_2) Z_{l\dr}(\t_1)+Z_{l\dr}(\t_2) Z_{k+1,\dq}(\t_1)\big)\Big]\,,
\eal
while if $l=k\ge 2$ then
\bal\la{zzk1drkdq}
Z_{k+1,\dr}(\t_1) Z_{k\dq}(\t_2) &= Z_{k+1,\dr}(\t_1)(\chi_{k-1}^-Z_{l-1,\dq}(\t_2) -Z_{k-1,\dq}(\t_2)\chi_{k-1}^- ) \\
&= \chi_{k-1}^-Z_{k+1,\dr}(\t_1)Z_{k-1,\dq}(\t_2)  -Z_{k+1,\dr}(\t_1)Z_{k-1,\dq}(\t_2) \chi_{k-1}^-  
\\
&=\Big[\chi_{k-1}^-\,,\, S(\t_{12})\Big[s_{12}^2 Z_{k+1,\dr}(\t_2) Z_{k-1,\dq}(\t_1)+s_{21}^2 Z_{k-1,\dq}(\t_2)Z_{k+1,\dr}(\t_1) \\
&\qquad+s_{12}s_{21}\big(Z_{k+1,\dq}(\t_2) Z_{k-1,\dr}(\t_1)+Z_{k-1,\dr}(\t_2) Z_{k+1,\dq}(\t_1)\big)\Big]\Big]\\
&=S(\t_{12})\Big[s_{12}^2 Z_{k+1,\dr}(\t_2) Z_{k\dq}(\t_1)+s_{21}^2 Z_{k\dq}(\t_2)Z_{k+1,\dr}(\t_1) \\
&\qquad+s_{12}s_{21}\big(Z_{k+1,\dq}(\t_2) Z_{k\dr}(\t_1)+Z_{k\dr}(\t_2) Z_{k+1,\dq}(\t_1)\big)\Big]\,.
\eal
Obviously, the consideration of the other indices follows the same line. 
Thus, it is sufficient  to prove that $Z_{1\d1}$, $Z_{2\d1}$, $Z_{1\dt}$, $Z_{2\dt}$ satisfy the ZF algebra.

%%%%%%%%%%%%%%%%
 \subsection{Integration contours and the ZF operators $Z_{2\d1}$, $Z_{1\dt}$, $Z_{2\dt}$}
 
 To simplify the computations it is often convenient to choose a particular ordering of integration contours in $Z_{m\dn}(\t)$ operators. In what follows in each $Z_{m\dn}(\t)$ all the integration contours $C_{k\dr}$ and $C_{\dk r}$ are shifted below $\t^{--}\equiv \t-2\pi i/N$. Then, the contours are arranged as follows
 \bal
 Z_{m\dn}(\t):\quad \t^{--}/C_{1\dr}/C_{\d1 r}/C_{2\dq}/C_{\dt q}/\cdots / C_{m-1,\ds}/C_{\dm-\d1, s}/\cdots /C_{\dn-\d1, w}\,,\quad m\le \dn\,,\\
  Z_{m\dn}(\t):\quad \t^{--}/C_{1\dr}/C_{\d1 r}/C_{2\dq}/C_{\dt q}/\cdots / C_{n-1,\ds}/C_{\dn-\d1, s}/\cdots /C_{m-1, \dw}\,,\quad m> \dn\,,
 \eal
 where $\dn$ and $n$ are identified, and the notation $C_{1\dr}/C_{\d1 r}$ means that the contour $C_{1\dr}$ is a bit above $C_{\d1 r}$. Then, the contours $C_{k\dr}$ and $C_{k+1,\dr}$  are separated by $2\pi i/N$. Finally, for all $k, \dk, \dr,\dq, r,q$ one takes $C_{k\dr}=C_{k\dq}$ and $C_{\dk r}=C_{\dk q}$.  This choice makes convenient to use the specific path to any ZF operator, see \eqref{path}.
 \bal\la{path}
\begin{array}{ccccccc}

 \vspace{0.15cm}
 
 Z_{1\d1}&\stackrel{\chi_1^-}{\longrightarrow} & Z_{2\d1} &\stackrel{\chi_2^-}{\longrightarrow} & Z_{3\d1} &\stackrel{\chi_3^-}{\longrightarrow} & Z_{4\d1}    \\ 
 
 \vspace{0.15cm}
 
{}_{\dchi_\d1^-}\big\downarrow ~ &   & {}_{\dchi_\d1^-}\big\downarrow &&&&\\

 \vspace{0.15cm}
 
 Z_{1\dt}&   & Z_{2\dt}&\stackrel{\chi_2^-}{\longrightarrow} & Z_{3\dt} &\stackrel{\chi_3^-}{\longrightarrow} & Z_{4\dt}     \\
 
  \vspace{0.15cm}
  
 {}_{\dchi_\dt^-}\big\downarrow ~ &   & {}_{\dchi_\dt^-}\big\downarrow &&{}_{\dchi_\dt^-}\big\downarrow&& \\
 
  \vspace{0.15cm}
  
 Z_{1\dth}& &Z_{2\dth} &&Z_{3\dth} &\stackrel{\chi_3^-}{\longrightarrow} &Z_{4\dth}     \\
  
   \vspace{0.15cm}
   
  {}_{\dchi_\dth^-}\big\downarrow ~ &   & {}_{\dchi_\dth^-}\big\downarrow &&{}_{\dchi_\dth^-}\big\downarrow&&{}_{\dchi_\dth^-}\big\downarrow\\
  
   \vspace{0.15cm}
   
  Z_{1\df}& &  Z_{2\df}&  &  Z_{3\df}& &  Z_{4\df} 
\end{array}
\eal
It is also particularly useful to analyse the highest weight   
 $r$-particle bound state vertex operators  discussed in section \ref{boundstates}.
 
  \bigskip
 
All integration contours in the formulae below are canonically ordered, and the notation  $R_{A|B}\equiv i\, r_{A|B}$  is used where $r_{A|B}$ is the residue of $g_{A|B}$. Since any of the Green's functions has no more than one pole there is no ambiguity in the definition.

\medskip

The derivation of the ZF operators $Z_{2\d1}$, $Z_{1\dt}$, $Z_{2\dt}$ is given in appendix \ref{Z12op}. 

One gets $Z_{2\d1}$ and $Z_{1\dt}$
\bal\la{Z2d1}
Z_{2\d1}(\t)&=Z_{2\d1}^{(1)}(\t)+Z_{2\d1}^{(2)}(\t) \\
& =  \r(\t)\int_{\t^{--}/C_{1\dr}} \,c_{1\dr}\,g^a_{0|1\dr}\,e^{i\p_0+i\p_{1\dr}} +  \r(\t)c_{1\d1}(\t)\,R_{0|1\d1}\,e^{i\p_0+i\p_{1\d1}(\t)}\,,
\eal
\bal\la{Z1d2}
Z_{1\dt}(\t)&=Z_{1\dt}^{(1)}(\t)+Z_{1\dt}^{(2)}(\t)\\
&=  \r(\t)\int_{\t^{--}/C_{\d1 r}}\,c_{\d1 r}\,g^a_{0|\d1 r}\,e^{i\p_0+i\p_{\d1 r}} +  \r(\t)c_{\d1 1}(\t)R_{0|\d1 1}\,e^{i\p_0(\t)+i\dphi_{\d1 1}(\t)}\,.
\eal
Note that there is no pole at $\t_{12} = \un_2$ in the products $Z_{1\d1}(\t_1)Z_{2\d1}(\t_2)$ and $Z_{1\d1}(\t_1)Z_{1\dt}(\t_2)$
because, e.g. $R_{0|\d1 r}=\de_{r1}R_{0|\d1 1}$ but $g_{1\d1|0}=1$.

The operator $Z_{2\dt}$ is given by 
\bal\la{Z2d2}
Z_{2\dt}(\t) &=\sum_{a=1}^6Z_{2\dt}^{(a)}(\t) \,, 
\eal
\bal
Z_{2\dt}^{(1)}(\t)&= \r(\t)\int_{\t^{--}/C_{{1\dr}}/C_{{\d1 q}}}c_{1\dr} \,c_{\d1 q}\,g_{0|1\dr}^a\,g_{0|\d1 q}^a\,
g_{\d1 q|1\dr}\,e^{i\p_0+i\phi_{1\dr}+i\dphi_{\d1 q}}\,,
\eal
\bal
Z_{2\dt}^{(2)}(\t)&= \r(\t)\int_{\t^{--}/C_{{\d1 q}}} \,c_{1\d1}(\t) \,c_{\d1 q}\,R_{0|1\d1}\,g^a_{0|\d1 q}\, g_{\d1 q|1\d1}(\a_{\d1 q}-\t)\,e^{i\p_0+i\phi_{1\d1}(\t)+i\dphi_{\d1 q}}\,,
\eal
\bal
Z_{2\dt}^{(3)}(\t)&= \r(\t)\int_{\t^{--}/C_{{1\dr}}} \,c_{1\dr} \,c_{\d1 1}(\t)\,g_{0|1\dr}^a\,R_{0|\d1 1}\, g_{1\dr|\d1 1}(\a_{1\dr}-\t)\,e^{i\p_0+i\phi_{1\dr}+i\dphi_{\d1 1}(\t)}\,,\eal
\bal
Z_{2\dt}^{(4)}(\t)&= \r(\t) \int_{\t^{--}/C_{{1\dr}}} \,c_{1\dr} \,c_{\d1 q}(\a_{1\dr})\,g_{0|1\dr}^a\,
g_{0|\d1 q}(\t-\a_{1\dr})\, R_{1\dr|\d1 q}\,e^{i\p_0+i\phi_{1\dr}+i\dphi_{\d1 q}(\a_{1\dr})}\,,
\eal
\bal
Z_{2\dt}^{(5)}(\t)&= \r(\t)\,c_{1\d1}(\t) \,c_{\d1 1}(\t)\,R_{0|1\d1}\,R_{0|\d1 1}\,e^{i\p_0+i\phi_{1\d1}(\t)+i\dphi_{\d1 1}(\t)}\,,
\eal
\bal
Z_{2\dt}^{(6)}(\t)&=  \r(\t)\,c_{1\d1}(\t) \,c_{\d1 2}(\t)\,R_{0|1\d1}\, R_{1\d1|\d1 2}\,e^{i\p_0+i\phi_{1\d1}(\t)+i\dphi_{\d1 2}(\t)}
\,.
\eal

%%%%%%%%%%%%%%%%
\subsection{The ZF algebra for $Z_{1\d1}$, $Z_{2\d1}$, $Z_{1\dt}$, $Z_{2\dt}$}

It is not difficult to show that all ZF relations for $Z_{1\d1}$, $Z_{2\d1}$, $Z_{1\dt}$, $Z_{2\dt}$ follow from the commutativity of the left and right charges 
\bal
\chi^-_1 \,\dot\chi^-_\d1=\dot\chi^-_\d1\,\chi^-_1  \,,
\eal
the commutativity of the charges with the following ZF operators
\bal
\chi^-_1 \,Z_{2\dk}(\t)=Z_{2\dk}(\t)\,\chi^-_1  \,,\quad
\dot\chi^-_\d1 \,Z_{k\dt}(\t)=Z_{k\dt}(\t)\,\dot\chi^-_\d1\,,\quad \dk = \d1,\dt\,,\quad k = 1,2\,,
\eal
and 
the basic relations
\bal
Z_{1\d1}(\t_1) Z_{1\d1}(\t_2) =S(\t_{12})Z_{1\d1}(\t_2) Z_{1\d1}(\t_1)\,,
\eal
\bal
Z_{1\d1}(\t_1) Z_{2\d1}(\t_2) =S(\t_{12})\left[s_{12} Z_{1\d1}(\t_2) Z_{2\d1}(\t_1)+s_{21}Z_{2\d1}(\t_2) Z_{1\d1}(\t_1)\right]\,,
\eal
\bal
Z_{1\d1}(\t_1) Z_{1\dt}(\t_2) =S(\t_{12})\left[s_{12} Z_{1\d1}(\t_2) Z_{1\dt}(\t_1)+s_{21}Z_{1\dt}(\t_2) Z_{1\d1}(\t_1)\right]\,,
\eal
\bal\la{zz1d12d2}
Z_{1\d1}(\t_1) Z_{2\dt}(\t_2) =S(\t_{12})\Big[&s_{12}^2 Z_{1\d1}(\t_2) Z_{2\dt}(\t_1)+s_{21}^2 Z_{2\dt}(\t_2)Z_{1\d1}(\t_1) \\
&+s_{12}s_{21}\big(Z_{2\d1}(\t_2) Z_{1\dt}(\t_1)+Z_{1\dt}(\t_2) Z_{2\d1}(\t_1)\big)\Big]\,.
\eal
It is easy to verify all these relations by using the formulae for the ZF operators from the previous subsection.  As was mentioned before, the commutativity of charges follows from the relations \eqref{fieldrel1} and \eqref{fieldrel2}. As an example, let us prove the commutativity of the charge $\chi^-_1$ with $Z_{2\d1}$. 

One has (dropping $\r(\t)$)
\bal
\chi^-_1Z_{2\d1}^{(1)}(\t) &\sim \int_{C_{\a_\dr}/\t/\t^{--}/C_{\a_\dq}} \,c_{1\dr} \,c_{1 \dq}\, g_{0|1 \dq}^a \, g_{0|1\dr}g_{1 \dq|1\dr} 
 \,e^{i\p_0+i\phi_{1\dr}+i\phi_{1 \dq}}\\
 &= \int_{\t^{--}/C_{\a_\dr}=C_{\a_\dq}} \,c_{1\dr} \,c_{1 \dq}\, g_{0|1 \dq}^a \, g_{0|1\dr}g_{1 \dq|1\dr} 
 \,e^{i\p_0+i\phi_{1\dr}+i\phi_{1 \dq}}\\
 &+ \int_{\t^{--}/C_{\a_\dq}} \,c_{1\dr}(\t) \,c_{1 \dq}\, g_{0|1 \dq}^a \, R_{0|1\dr}g_{1 \dq|1\dr}(\a_{1\dq}-\t) 
 \,e^{i\p_0+i\phi_{1\dr}(\t)+i\phi_{1 \dq}}\,,
\eal
\bal
\chi^-_1Z_{2\d1}^{(2)}(\t) &\sim \int_{C_{\a_\dr}/\t} \,c_{1\dr} \,c_{1 \dq}(\t)\, R_{0|1 \dq} \, g_{0|1\dr}g_{1 \dq|1\dr}(\t-\a_{1\dr}) 
 \,e^{i\p_0+i\phi_{1\dr}+i\phi_{1 \dq}(\t)}\\
 &= \int_{\t^{--}/C_{\a_\dr}} \,c_{1\dr} \,c_{1 \dq}(\t)\, R_{0|1 \dq} \, g_{0|1\dr}g_{1 \dq|1\dr}(\t-\a_{1\dr}) 
 \,e^{i\p_0+i\phi_{1\dr}+i\phi_{1 \dq}(\t)}\,,
\eal
where one used that $g_{1 \d1|1\d1}(0) =0$.
One also has
\bal
Z_{2\d1}^{(1)}(\t)\chi^-_1 &\sim \int_{\t^{--}/C_{\a_\dr}=C_{\a_\dq}}c_{1\dr}(\a_\dr) \,c_{1\dr} \,c_{1 \dq}\, g_{0|1 \dq}^a \, g_{1\dr|0}g_{1 \dr|1\dq} 
 \,e^{i\p_0+i\phi_{1\dr}+i\phi_{1 \dq}}\,,
\eal
\bal
Z_{2\d1}^{(2)}(\t)\chi^-_1 &\sim \int_{\t^{--}/C_{\a_\dr}} \,c_{1\dr} \,c_{1 \dq}(\t)\, R_{0|1 \dq} \, g_{1\dr|0}g_{1 \dr|1\dq}(\a_{1\dr}-\t) 
 \,e^{i\p_0+i\phi_{1\dr}+i\phi_{1 \dq}(\t)}\,,
\eal
where there is no need to shift the integration contour $C_{\a_\dr}$ . It is easy to check that the integrands of the double integrals are equal to each other after  symmetrisation with respect to $\dr$, $\dq$, and that the integrands of the single integrals also add up to 0.

%%%%%%%%%%%%%%%%
\section{Bound states}\la{boundstates}

The  ordering of integration contours discussed in the previous section is particularly useful to analyse the highest weight   
 $r$-particle bound state vertex operators\footnote{In this section the dotted and undotted indices are identified that is if $r$ is equal to a number then $\dr$ is equal to the same number with a dot, e.g. if $r=3$ then $\dr =\dth$. }
 \bal
 Z_{12...r,\d1\dt...\dr}(\t)=  \lim_{\eps_{i+1,i}\to 0}\prod_{j=2}^{r}(i\eps_{j+1,j} )\,Z_{1\d1}\big(\t_1^\eps\big)Z_{2\dt}\big(\t_2^\eps\big)\cdots Z_{r\dr}\big(\t_r^\eps\big)\,,\quad \t_j^\eps\equiv \t +\un_{r-2j+1}+\eps_j\,,
\eal
where  $\un_k\equiv {\pi\ov N}i\,k$ is purely imaginary,  $\eps_1=0$  and all $\eps_{jk}\equiv \eps_{j}-\eps_{k}$ do not vanish until one takes the limits. 

Similarly to bound states of the chiral GN model \cite{BF13}, any rank-$r$ bound state  vertex operator is generated from the  highest weight vertex operator $Z_{12...r,\d1\dt...\dr}$ by acting on it with the lowering symmetry operators. 

The highest weight multi-particle bound state vertex operators $Z_{12...r,\d1\dt...\dr}$ can be found from the terms in ZF operators which have no integration after the integration contours have been reduced to the canonical form.

%%%%%%%%%%%%%%%%%%%%%%%%%%%%%%%%%%%%%%%
\subsection*{Highest weight 2-particle bound state}

It is easy to see that the products $Z_{1\d1}(\t_1)Z_{2\dt}(\t_2)$, $Z_{2\dt}(\t_1)Z_{1\d1}(\t_2)$, $Z_{2\d1}(\t_1)Z_{1\dt}(\t_2)$ and $Z_{1\dt}(\t_1)Z_{2\d1}(\t_2)$ have a pole at $\t_{12}=2\pi i/N$ which means that for $N\ge 3$ these operators form a bound state, while for $N=2$ the pole shows that $Z_{1\d1}$, $Z_{2\dt}$,  and $Z_{1\dt}$, $Z_{2\d1}$ are particle-antiparticle pairs.

To find $Z_{12,\d1\dt}$ let us consider $Z_{1\d1}(\t_1)Z_{2\dt}(\t_2)$ in the limit $\t_{12}\to \un_2$. Since the canonically-ordered integration contours in $Z_{2\dt}$ run below all poles and $g_{1\d1|0}=g_{\d1 1|0}=1$,  only the last term in \eqref{Z2d2} produces the bound state pole. 
 Introducing the notation
\bal \la{U2d2n}
U_{2\dt}(\t)&=\cN_{2\dt}(\t)\,e^{i\p_0(\t)+i\phi_{1\d1}(\t)+i\dphi_{\d1 2}(\t)}\,,\quad \cN_{2\dt}(\t)= \r(\t)c_{1\d1}(\t) \,c_{\d1 2}(\t)\,R_{0|1\d1}\, R_{1\d1 |\d1 2}\,,
\eal
one finds
\bal
&Z_{1\d1}(\t_1)U_{2\dt}(\t_2)=\r(\t_1)\cN_{2\dt}(\t_2) \,g_{00}(\t_{21})\,g_{\d1 2|0}(\t_{21})\,e^{i\p_0(\t_1)+i\p_0(\t_2)+i\phi_{1\d1}(\t_2)+i\dphi_{\d1 2}(\t_2)}\\
&=-{g_{00}(-\un_2)r_{\d1 2|0}\ov \t_{12}-\un_2}\,\r(\t_1)\cN_{2\dt}(\t_2) \,e^{i\p_0(\t_1)+i\p_0(\t_2)+i\phi_{1\d1}(\t_2)+i\dphi_{\d1 2}(\t_2)}+\cO(1)\\
&=  {i\ov\t_{12}-\un_2}  Z_{12,\d1\dt}(\t) + \cO(1)\,,\quad \t_1\to\t^+\equiv\t+\un_1\,,\ \t_2\to\t^-\equiv\t-\un_1\,.
\eal
Here the highest weight $2$-particle bound state vertex operator $Z_{(2)}\equiv Z_{12,\d1\dt}$ is given by
\bal
 Z_{(2)}(\t) &=  \cN_{(2)}(\t) \,e^{i\p_0(\t^+)+i\p_0(\t^-)+i\phi_{1\d1}(\t^-)+i\dphi_{\d1 2}(\t^-)}\,,\\
   \cN_{(2)}(\t) &= i\,{g_{00}(-\un_{2})r_{\d1 2|0}}\,\r(\t^+)\cN_{2\dt}(\t^-) \,.
\eal
Thus, up to a normalisation the vertex operator 
\bal\la{v12d1d2}
V_{(2)}(\t)=e^{i\p_{(2)}(\t)}\,,\quad \p_{(2)}(\t)=\p_0(\t^+) +\p_0(\t^-)+\phi_{1\d1}(\t^-)+\dphi_{\d1 2}(\t^-)\,,
\eal
creates  the highest weight 2-particle bound state.
 All the other 2-particle bound states are obtained from $Z_{(2)}$ by acting on it with the charges $\chi_k^-\,,\, \dot\chi_\dk^-$. It is easy to check that the fields $\p_{1\dr}$ and $\dot\p_{\d1 r}$ have trivial Green's functions with the $2$-particle bound state free field $\p_{(2)}$\footnote{Obviously, $\p_{k\dr}$ and $\dot\p_{\dk r}$ for $k\ge3$ also have trivial Green's functions with $\p_{(2)}$.}
\bal \bra\p_{1\dr}(\a)\p_{(2)}(\t)\ket =\bra\dot\p_{\d1 r}(\a)\p_{(2)}(\t)\ket =\bra\p_{(2)}(\t)\p_{1\dr}(\a)\ket =\bra\p_{(2)}(\t)\dot\p_{\d1 r}(\a)\ket =1\,.
\eal
Thus, the charges $\chi_1^-\,,\, \dot\chi_\d1^-$ commute with $Z_{(2)}$
\bal 
\big[\chi_1^-, Z_{(2)}(\t)\big] =\big[\dot\chi_\d1^-, Z_{(2)}(\t)\big] =0\,,
\eal
which is a necessary condition for $Z_{(2)}$ to be the highest weight 2-particle bound state.

\medskip

For $N=2$ the operators $Z_{1\d1}$ and $Z_{2\dt}$ form a particle-antiparticle pair, and the operator $Z_{(2)}(\t)$ must be equal to $-C_{1\d1,2\dt}=-1$ where $C_{k\dl,m\dn}=\e_{km}\e_{\dl\dn}$ is the charge conjugation matrix. Thus, one finds the following relation between the free fields
\bal 
\p_0(\t^{++})+\p_0(\t)+\phi_{1\d1}(\t)+\dphi_{\d1 2}(\t) =0\,,
\eal
which together with 
\bal
 \phi_{1\d1}(\t)+\dphi_{\d1 2}(\t)=\phi_{1\dt}(\t)+\dphi_{\d1 1}(\t) \,,
\eal
reduces the number of independent fields to 3.
Then,  the normalisation condition $   \cN_{(2)}(\t)=-1$ leads to
\bal
 {g_{00}(-\un_{2})\,R_{\d1 2|0}}\,R_{0|1\d1}\, R_{1\d1 |\d1 2}\,\r(\t^+)\r(\t^-)c_{1\d1}(\t^-) \,c_{\d1 2}(\t^-)=-1\,.
 \eal
Taking into account that 
for $N=2$
\bal
{g_{00}(-\un_{2}) ={2\ov\pi}\,,\quad  R_{\d1 2|0}}=R_{0|1\d1}= R_{1\d1 |\d1 2}=\pi\,,
\eal
 one gets
\bal\la{norm2}
2\pi^2\r(\t^{++}) \r(\t)c_{1\d1}(\t) \,c_{\d1 2}(\t)  =-1\,.
\eal
Now, according to \eqref{rhocc}
\bal
\r(\t)=e^{i\varkappa\,\t}\tilde\r\,,\quad c_{k\dr}(\t)=e^{i\vk_{k} \t}\tilde c_{k\dr}\,,\quad c_{\dk r}(\t)=e^{i\vk_{\dk} \t}\tilde c_{\dk r}\,,
\eal
and therefore the equation \eqref{norm2} splits into the following two relations
\bal
2\vk + \vk_1+\vk_\d1&=0\,,\quad 2\pi^2\,e^{-{\pi}\vk}\,  \tilde\r^2\, \tilde c_{1\d1}\,\tilde c_{\d1 2}=-1\,.
\eal
These two relations together with the equation \eqref{fieldrel2} provide 3 constraints on the 8 parameters. In addition, it is clear that for $N=2$ the extended ZF algebra is invariant under the following scaling transformations
\bal
Z_{1\d1}\to \s_{1\d1} Z_{1\d1}\,,\quad \chi_{1}\to \s_{1} \chi_{1}\,,\quad \dchi_{\d1}\to {1\ov\s_{1\d1}^2\s_1} \dchi_{\d1}\,,
\eal
where $\s_{1\d1}$ and $\s_1$ are arbitrary constants.
Thus, one can set $\tilde c_{1\d1}$ and $\tilde c_{\d1 2}$ to any values, and therefore the number of independent parameters leading to physically distinct results is 3 which agrees with \cite{FL}.
For example, choosing $\tilde c_{1\d1}=e^{-{\pi}\vk_1/2}$, $\tilde c_{\d1 2}=e^{-{\pi}\vk_\d1/2}$, one gets
$\tilde\r={i\ov \sqrt2\pi}$,  and the following two constraints on the remaining five parameters
\bal
2\vk + \vk_1+\vk_\d1&=0\,,\quad e^{-{\pi}\vk}\,\tilde c_{1\dt}\,\tilde c_{\d1 1}=1\,.
\eal

%%%%%%%%%%%%%%%%
\subsection*{Highest weight 3-particle bound state}

As was mentioned above, the $2$-particle bound state field $\p_{(2)}$ has trivial Green's functions with
  $\p_{k\dr}$, $\dot\p_{\dk r}$ for $k\neq 2$, $\dk\neq\dt$. It is easy to find that it has the following Green's functions with $\p_0$, $\p_{2\dq}$, $\dot\p_{\dt q}$
\bal
g_{0(2)}(\t )=g_{(2)0}(\t )=\frac{\Gamma \left(\frac{i \t}{2 \pi }-\frac{3}{2 N}+1\right) \Gamma \left(\frac{i
   \t}{2 \pi }+\frac{3}{2 N}\right)}{\Gamma \left(\frac{i \t}{2 \pi }-\frac{1}{2
   N}+1\right) \Gamma \left(\frac{i \t}{2\pi
   }+\frac{1}{2N}\right)}\,,
\eal
\bal
& g_{2\dq|(2)}(\t )=g_{\dt q|(2)}(\t)={\t+ \un_{1}\ov\t+\un_3}\,,\quad g_{(2)|2\dq}(\t )=g_{(2)|\dt q}(\t )=1\,,\quad q\ge 3\,,\ \dq \ge \dth\,,
\eal
\bal
& g_{2\dq|(2)}(\t )=g_{\dt q|(2)}(\t )=1\,,\quad  g_{(2)|2\dq}(\t )=g_{(2)|\dt q}(\t )={\t-\un_3\ov\t- \un_{1}}\,,\quad q\le 2\,,\ \dq \le \dt\,.
\eal
Thus, one can get a pole in $Z_{12,\d1\dt}(\t_1)Z_{3\dth}(\t_2)$ at $\t_{12}=-\un_3$ only if  $Z_{3\dth}(\t_2)$ contains a term with $\p_{2\dth}$ or $\p_{\dt 3}$. The terms  in $Z_{3\dth}$ without integration are obtained from
\bal
 \dchi_{\dt}^-\chi_2^- U_{2\dt} =  \cN_{2\dt}(\t)\int c_{2\dr}c_{\dt q}\,g_{1\d1|2\dr}\,g_{\d1 2|2\dr}\,g_{1\d1|\dt q}\,g_{\d1 2|\dt q} \,g_{2\dr|\dt q}\,e^{i\p_0(\t)+i\phi_{1\d1}(\t)+i\dphi_{\d1 2}(\t)+i\phi_{2\dr}+i\dphi_{\dt q}} \,.
\eal
If $\dr = \dth$ then the function $g_{1\d1|2\dr}\,g_{\d1 2|2\dr}$ has no pole, and taking the contour $C_{2\dr}$ below $\t^{--}$ would not produce any term without integration. Thus, $q=3$ and therefore $\dr=\dt$, and the term of interest   is
\bal\la{U3d3n}
 U_{3\dth}(\t) &=  \cN_{3\dth}(\t)\,e^{i\p_0(\t)+i\phi_{1\d1}(\t)+i\dphi_{\d1 2}(\t)+i\phi_{2\dt}(\t)+i\dphi_{\dt 3}(\t)}\,,\\
   \cN_{3\dth}(\t)&= \cN_{2\dt}(\t)c_{2\dt}(\t)c_{\dt 3}(\t)\,R_{\d1 2|2\dt}\,R_{2\dt|\dt 3} \,.
\eal
Then, the highest weight 3-particle bound state vertex operator $Z_{(3)}\equiv Z_{123,\d1\dt\dth}$ appears in the residue of the product $Z_{(2)}(\t_1)U_{3\dth}(\t_2)$ at $\t_{12}=3\pi i/N$
\bal
&Z_{(3)}(\t) =\lim_{\eps\to 0}\,i\eps\,Z_{(2)}(\t+\un_1) U_{3\dth}(\t-\un_2+\eps)\\
&=\cN_{(3)}(\t)\, e^{i\p_{(2)}(\t^{+})+i\p_0(\t^{--})+i\phi_{1\d1}(\t^{--})+i\dphi_{\d1 2}(\t^{--})+i\phi_{2\dt}(\t^{--})+i\dphi_{\dt 3}(\t^{--})}\\
&\cN_{(3)}(\t) 
=i\,g_{0(2)}(-\un_3) r_{2\dth|(2)}\,\cN_{(2)}(\t+\un_1) \cN_{3\dth}(\t-\un_2)
\,.
\eal
Thus,  up to a normalisation the vertex operator
\bal
V_{(3)}(\t)&=e^{i\p_{(3)}(\t)}\,,\\
\p_{(3)}(\t)&=\p_0(\t^{++})+\p_0(\t)+\p_0(\t^{--})+\phi_{1\d1}(\t)+\phi_{1\d1}(\t^{--})\\
&+\dphi_{\d12}(\t)+\dphi_{\d1 2}(\t^{--})+\phi_{2\dt}(\t^{--})+\dphi_{\dt 3}(\t^{--})
\eal
creates the highest weight 3-particle bound state for $N>3$, while for $N=3$ the operator $Z_{(2)}$  creates the anti-particle of $Z_{3\dth}$, and therefore $V_{(3)}$ must be equal to 1
\bal
V_{(3)}(\t)=1\,, \quad N=3\,,
\eal
leading to the relation
\bal
\p_0(\t^{++})&+\p_0(\t)+\p_0(\t^{--})+\phi_{1\d1}(\t)+\phi_{1\d1}(\t^{--})\\
&+\dphi_{\d12}(\t)+\dphi_{\d1 2}(\t^{--})+\phi_{2\dt}(\t^{--})+\dphi_{\dt 3}(\t^{--})=0\,, \quad N=3\,.
\eal
Then,  the normalisation condition $   \cN_{(3)}(\t)=-1$ leads again  to two equations on the 
parameters $\vk$, $\vk_k$, $\vk_\dk$, $\tilde c_{k\dr}$, $\tilde c_{\dk r}$ which will be discussed later for the general $N$ case.

%%%%%%%%%%%%%%%%
\subsection*{Highest weight 4-particle bound state}

It is easy to verify that the $3$-particle bound state field $\p_{(3)}$ has trivial Green's functions with
 $\p_{k\dr}$, $\p_{\dk r}$ for $k\neq 3$, $\dk\neq\dth$, and it has the following Green's functions with $\p_0$, $\p_{3\dq}$, $\p_{\dth q}$
\bal
g_{0(3)}(\t )=g_{(3)0}(\t )=\frac{\Gamma \left(\frac{i \t}{2 \pi }-\frac{2}{ N}+1\right) \Gamma \left(\frac{i
   \t}{2 \pi }+\frac{2}{ N}\right)}{\Gamma \left(\frac{i \t}{2 \pi }-\frac{1}{
   N}+1\right) \Gamma \left(\frac{i \t}{2\pi
   }+\frac{1}{N}\right)}\,,
\eal
\bal
& g_{3\dq|(3)}(\t )=g_{\dth q|(3)}(\t)={\t+ \un_{2}\ov\t+\un_4}\,,\quad g_{(3)|3\dq}(\t )=g_{(3)|\dth q}(\t )=1\,,\quad q\ge 4\,,\ \dq \ge \df\,,
\eal
\bal
& g_{3\dq|(3)}(\t )=g_{\dth q|(3)}(\t )=1\,,\quad  g_{(3)|3\dq}(\t )=g_{(3)|\dth q}(\t )={\t-\un_4\ov\t- \un_{2}}\,,\quad q\le 3\,,\ \dq \le \dth\,.
\eal
Thus, one can get a pole in $Z_{123,\d1\dt\dth}(\t_1)Z_{4\df}(\t_2)$ at $\t_{12}=-\un_4$ only if  $Z_{4\df}(\t_2)$ contains a term with $\p_{3\df}$ or $\p_{\dth 4}$. The terms  in $Z_{4\df}$ without integration are obtained from
\bal
 \dchi_{\dth}^-\chi_3^- U_{3\dth} =  \cN_{3\dth}(\t)\int & c_{3\dr}c_{\dth q}\,g_{2\dt|3\dr}\,g_{\dt 3|3\dr}\,g_{2\dt|\dth q}\,g_{\dt 3|\dth q} \,g_{3\dr|\dth q}\,\\
 &\times e^{i\p_0(\t)+i\phi_{1\d1}(\t)+i\dphi_{\d1 2}(\t)+i\phi_{2\dt}(\t)+i\dphi_{\dt 3}(\t)+i\phi_{3\dr}+i\dphi_{\dth q}} \,,
\eal
and the term without integration is obtained for $q=4$ and $\dr=\dth$
\bal\la{U4d4n}
 U_{4\df}(\t) &=  \cN_{4\df}(\t)\,e^{i\p_0(\t)+i\phi_{1\d1}(\t)+i\dphi_{\d1 2}(\t)+i\phi_{2\dt}(\t)+i\dphi_{\dt 3}(\t)+i\phi_{3\dth}(\t)+i\dphi_{\dth 4}(\t)} \,,\\
   \cN_{4\df}(\t)&= \cN_{3\dth}(\t)c_{3\dth}(\t)c_{\dth 4}(\t)\,R_{\dt 3|3\dth}\,R_{3\dth|\dth 4} \,.
\eal
Then, the highest weight 4-particle bound state vertex operator $Z_{(4)}\equiv Z_{1234,\d1\dt\dth\df}$ appears in the residue of the product $Z_{(3)}(\t_1)U_{4\df}(\t_2)$ at $\t_{12}=4\pi i/N$
\bal
Z_{(4)}(\t) &=\lim_{\eps\to 0}\,i\eps\,\cZ_{(3)}(\t+\un_1) U_{4\df}(\t-\un_3+\eps)=\cN_{(4)}(\t) \, e^{i\p_{(4)}(\t)}\,,\\
\cN_{(4)}(\t) 
&=i\,g_{0(3)}(-\un_4) r_{3\df|(3)}\,\cN_{(3)}(\t+\un_1) \cN_{4\df}(\t-\un_3)\,,
\\
\p_{(4)}(\t)&= \sum_{k=-{3\ov 2}}^{{3\ov 2}}\p_0(\t+\un_{2k})+\sum_{n=1}^{3}\sum_{k=-{3\ov 2}}^{{3\ov 2}-n}\big(\p_{n\dn}(\t+\un_{2k})+\p_{\dn, n+1}(\t+\un_{2k})\big)\,.
\eal
Thus, up to a normalisation the vertex operator $V_{(4)}(\t)=e^{i\p_{(4)}(\t)}$ creates the highest weight 4-particle bound state for $N>4$, while for $N=4$ the operator $Z_{(3)}$  creates the anti-particle of $Z_{4\df}$, and therefore $V_{(4)}$  satisfies
\bal
V_{(4)}(\t)=1\,, \quad N=4\,.
\eal
  Moreover, for any $N$ the $4$-particle bound state field $\p_{(4)}$ has trivial Green's functions with
 with $\p_{k\dr}$, $\p_{\dk r}$ for $k\neq 4$, $\dk\neq\df$.
  
  %%%%%%%%%%%%%%%%
\subsection*{Highest weight $r$-particle bound state}

The formulae above can be easily generalised, and the highest weight $r$-particle bound state vertex operator $Z_{(r)}\equiv Z_{12...r,\d1\dt...\dr}$ is given by
\bal\la{Zbr}
Z_{(r)}(\t) &=\lim_{\eps\to 0}\,i\eps\,\cZ_{(r-1)}(\t+\un_1) U_{r\dr}(\t-\un_{r-1}+\eps)=\cN_{(r)}(\t) \, e^{i\p_{(r)}(\t)}\,,
\\
\cN_{(r)}(\t) 
&=i\,g_{0(r-1)}(-\un_r) r_{r-1,\dr|(r-1)}\,\cN_{(r-1)}(\t+\un_1) \cN_{r\dr}(\t-\un_{r-1})\,,\\
   \cN_{r\dr}(\t)&= \cN_{r-1,\dr-\d1}(\t)c_{r-1,\dr-\d1}(\t)c_{\dr-\d1, r}(\t)\,R_{\dr-\dt, r-1|r-1,\dr-\d1}\,R_{r-1,\dr-\d1|\dr-\d1 r} \,,
\\
\p_{(r)}(\t) &=\sum_{k=-{r-1\ov 2}}^{{r-1\ov 2}}\p_0(\t+\un_{2k})+\sum_{n=1}^{r-1}\sum_{k=-{r-1\ov 2}}^{{r-1\ov 2}-n}\big(\p_{n\dn}(\t+\un_{2k})+\p_{\dn, n+1}(\t+\un_{2k})\big)\,.
\eal
The $r$-particle bound state field $\p_{(r)}$ commutes with $\p_{k\dr}$, $\p_{\dk r}$ for $k\neq r$, $\dk\neq\dr$, and it has the following Green's functions with $\p_0$, $\p_{r\dq}$, $\p_{\dr q}$
\bal
g_{0(r)}(\t )=g_{(r)0}(\t )&=\frac{\Gamma \left(\frac{i \t}{2 \pi }-\frac{r+1}{ 2N}+1\right) \Gamma \left(\frac{i
   \t}{2 \pi }+\frac{r+1}{ 2N}\right)}{\Gamma \left(\frac{i \t}{2 \pi }-\frac{r-1}{
   2N}+1\right) \Gamma \left(\frac{i \t}{2\pi
   }+\frac{r-1}{2N}\right)}\,,\\ 
   g_{0(r)}(-\un_{r+1}) &= \frac{r \Gamma \left(\frac{r+1}{N}\right)}{\Gamma \left(\frac{1}{N}\right) \Gamma
   \left(1+\frac{r}{N}\right)}\,,
\eal
\bal
g_{r\dq|(r)}(\t )=g_{\dr q|(r)}(\t )={\t+ \un_{r-1}\ov\t+ \un_{r+1}}\,,\quad g_{(r)|r\dq}(\t )=g_{(r)|\dr q}(\t )=1\,,\quad \dq > \dr\,,
\eal
\bal
g_{r\dq|(r)}(\t )=g_{\dr q|(r)}(\t )=1\,,\quad g_{(r)|r\dq}(\t )=g_{(r)|\dr q}(\t )={\t- \un_{r+1}\ov\t- \un_{r-1}}\,,\quad \dq\le \dr\,.
\eal
Explicit formulae expressing $\p_0$ and $\p_{(r)}$ in terms of the elementary fields  are collected in appendix \ref{solvconstr}. 

The normalisation functions $\cN_{(r)}(\t) $ can be found explicitly.  First, one expresses them in terms of  $\cN_{k\dk}$
\bal
\cN_{(r)}(\t) 
&=R_{0|1\d1}^{r-1} \prod_{k=1}^{r-1}g_{0(k)}(-\un_{k+1}) \prod_{k=1}^{r} \cN_{k\dk}(\t-\un_{2k-r-1})\,,
\eal
where $\cN_{1\d1}(\t)\equiv \r(\t)$, and one takes into account that the residues of all the functions in the formula \eqref{Zbr} are equal to $r_{0|1\d1}=-2\pi i/N$.
Next, one gets
\bal
\cN_{k\dk}(\t)
&=\r(\t) R_{0|1\d1}^{2(k-1)} \prod_{m=1}^{k-1}c_{m\dm}(\t)c_{\dm,m+1}(\t) \,,
\eal
\bal
\prod_{k=1}^{r-1}g_{0(k)}(-\un_{k+1})
&=\frac{\Gamma \left(\frac{r}{N}\right)}{\Gamma
   \left(\frac{1}{N}\right)\Gamma \left(1+\frac{1}{N}\right)^{r-1} }\,,
\eal
and therefore
\bal
\cN_{(r)}(\t) 
&=\frac{\Gamma \left(\frac{r}{N}\right)}{\Gamma
   \left(\frac{1}{N}\right)\Gamma \left(1+\frac{1}{N}\right)^{r-1} }R_{0|1\d1}^{r^2-1} \prod_{k=1}^{r}\r(\t-\un_{2k-r-1})\\
   &\times \prod_{k=1}^{r}\prod_{m=1}^{k-1}c_{m\dm}(\t-\un_{2k-r-1})c_{\dm,m+1}(\t-\un_{2k-r-1})\,.
\eal
Finally, taking into account \eqref{rhocc}, one gets
\bal\la{Nbr}
\cN_{(r)}(\t) 
&=\frac{\Gamma \left(\frac{r}{N}\right)}{\Gamma
   \left(\frac{1}{N}\right)\Gamma \left(1+\frac{1}{N}\right)^{r-1} }\left(\frac{2\pi}{N}\right)^{r^2-1}  \tilde\r^r e^{i r\vk \t}  \prod_{k=1}^{r-1}e^{i(r-k)(\vk_k+\vk_\dk)\t}\\
   &\times  \prod_{m=1}^{r-1}\left(e^{{\pi\ov N} m\vk_m} \tilde c_{m\dm}\right)^{r-m}
  \left( e^{{\pi\ov N} m\vk_\dm}\tilde c_{\dm,m+1}\right)^{r-m}\,.
\eal

The vertex operator $Z_{(N-1)}$ creates the anti-particle of $Z_{N\dN}$, and therefore 
$Z_{(N)}$ must be equal to $ -1$. This leads to the relation 
 $\p_{(N)} =0$,  which allows one to express $\p_0$ in terms of the other fields
\bal\la{popkr}
\sum_{k=-{N-1\ov 2}}^{{N-1\ov 2}}\p_0(\t+\un_{2k})+\sum_{n=1}^{N-1}\sum_{k=-{N-1\ov 2}}^{{N-1\ov 2}-n}\big(\p_{n\dn}(\t+\un_{2k})+\p_{\dn, n+1}(\t+\un_{2k})\big) = 0\,,
\eal
and to the normalisation condition
\bal
\cN_{(N)}(\t)  =-1.
\eal
It is important to stress that \eqref{popkr} leads to the following nontrivial crossing-type equation for  Green's function $g_{00}$
\bal\la{gogkr}
\prod_{k=-{N-1\ov 2}}^{{N-1\ov 2}}g_{00}(\t+\un_{2k})={\t+(1-{1\ov N})\pi i\ov \t-(1-{1\ov N})\pi i}\,,
\eal
which is indeed satisfied.

By using \eqref{Nbr}, the normalisation condition takes the form
\bal
\cN_{(N)}(\t) 
&=\frac{1}{\Gamma
   \left(\frac{1}{N}\right)\Gamma \left(1+\frac{1}{N}\right)^{N-1} }\left(\frac{2\pi}{N}\right)^{N^2-1}  \tilde\r^N e^{i N\vk \t}  \prod_{k=1}^{N-1}e^{i(N-k)(\vk_k+\vk_\dk)\t}\\
   &\times  \prod_{m=1}^{N-1}\left(e^{{\pi\ov N} m\vk_m} \tilde c_{m\dm}\right)^{N-m}
  \left( e^{{\pi\ov N} m\vk_\dm}\tilde c_{\dm,m+1}\right)^{N-m}=-1\,,
\eal
which splits into the following two equations
\bal\la{normconstraints}
&N\vk + \sum_{k=1}^{N-1}(N-k)(\vk_k+\vk_\dk)=0\,,\\
&
\frac{1}{\Gamma
   \left(\frac{1}{N}\right)\Gamma \left(1+\frac{1}{N}\right)^{N-1} }\left(\frac{2\pi}{N}\right)^{N^2-1} \tilde\r^N   \prod_{m=1}^{N-1}\left(e^{{\pi\ov N} m\vk_m} \tilde c_{m\dm}\right)^{N-m}
  \left( e^{{\pi\ov N} m\vk_\dm}\tilde c_{\dm,m+1}\right)^{N-m}=-1\,.
\eal
Adding these two constraints to the $(N-1)^2$ constraints on $\tilde c_{k\dm}$ and $\tilde c_{\dm k}$, one gets $N^2-2N+3$ constraints on the $2N^2$ parameters $\vk$, $\vk_k$, $\vk_\dk$, $\tilde\r$, $\tilde c_{k\dr}$, $\tilde c_{\dk r}$. 
The number of physically inequivalent parameters is however less then $N^2+2N-3$ because the extended ZF algebra is invariant under the following scaling transformations
\bal
Z_{1\d1}\to \s_{1\d1} Z_{1\d1}\,,\quad \chi_{k}\to \s_{k} \chi_{k}\,,\quad \dchi_{\dk}\to \s_\dk \dchi_{\d1}\,,
\eal
where $\s_{1\d1}$, $\s_k$ and $\s_\dk$ are  constants satisfying the constraint
\bal
\s_{1\d1}^N \prod_{m=1}^{N-1}\left(\s_{m}\s_{\dm}\right)^{N-m}=1\,.
\eal
Thus, one can choose the values of $2N-2$ parameters, and the number of independent parameters leading to physically distinct results is $N^2-1$ which is equal to the number of fields in the \sun2 PCF model. For example, choosing
\bal\la{tcm}
\tilde c_{m\dm}=e^{-{\pi\ov N} m\vk_m} \,,\quad \tilde c_{\dm,m+1}= e^{-{\pi\ov N} m\vk_\dm}\,,
\eal
one finds $\tilde\r$
\bal\la{rhov}
&
\tilde\r^N=-{\Gamma
   \left(\frac{1}{N}\right)\Gamma \left(1+\frac{1}{N}\right)^{N-1} }\left(\frac{N}{2\pi}\right)^{N^2-1} \,.
\eal

 \medskip

To conclude this section, it is worth mentioning that the normalisation condition $\cN_{(N)}=-1$ allows one to obtain the following expression for $\cN_{(N-1)}$
\bal
\cN_{(N-1)}(\t) 
&=\Gamma \left(1-\frac{1}{N}\right)\Gamma \left(1+\frac{1}{N}\right)\left(\frac{2\pi}{N}\right)^{1-2N}  {1\ov\tilde\r}e^{-i\vk \t}  e^{-i(N-1)(\vk_{N-1}+\vk_{\dN-\d1})\t } \\
   &\times  \prod_{m=1}^{r-1}\left(e^{{\pi\ov N} m\vk_m} \tilde c_{m\dm}\right)^{-1}
  \left( e^{{\pi\ov N} m\vk_\dm}\tilde c_{\dm,m+1}\right)^{-1}\,.
\eal

\section{Form factors} \la{form}

According to Lukyanov \cite{Lukyanov},  up to an overall normalisation constant
 the form factors  of the exponential operator $O$ corresponding to the constructed representation $\pi_O$ of the ZF algebra are 
 \be\la{Ffunc}
F^O_{\bI_1\ldots \bI_n}(\t_1,\ldots,\t_n)= \bra\bra Z_{\bI_n}(\t_n) \cdots Z_{\bI_1}(\t_1)  \ket\ket\,,
\ee
where  for any operator $W$ acting in $\pi_O$ the quantity $\bra\bra W \ket\ket$ is defined by
 \be\la{ffax3}
\bra\bra W \ket\ket= {\Tr_{\pi_{O}}\left[e^{2\pi i \,\bK}\,W\right]\ov\Tr_{\pi_{O}}\left[e^{2\pi i \,\bK}\right]}\,.
\ee
Due to the \sun2 symmetry the form factors do not vanish only for the states which do not carry charges with respect to the Cartan generators $P_k$ and $P_\dk$. 

It is clear that the form factors \eqref{Ffunc} are sums of multiple integrals with integrands of the form
\be\la{Rfunc}
 R_{\mu_1\ldots \mu_q}(\b_1,\ldots,\b_q)=\langle\langle V_{\mu_q}(\b_{q})\cdots V_{\mu_1}(\b_{1}) \rangle\rangle\,,
\ee
where the set $\{\b_1,\ldots,\b_q\}$ contains $\t_j\,$-rapidities. It is shown in appendix \ref{traces} that for any operator $W$ which is the product of free field exponents
\be
W = V_n(\t_n)\cdots V_1(\t_1)\,,\quad V_j(\t) =\, :e^{i\p_j(\t)}:\, =\, e^{i\p_j^+(\t)}\, e^{i\p_j^-(\t)}\,,\quad 
\ee
one obtains $\langle\langle W \rangle\rangle$ by applying the Wick theorem
\be
\langle\langle V_n(\t_n)\cdots V_1(\t_1) \rangle\rangle =\prod_{j=1}^nC_{V_j}\prod_{k<j}G_{V_kV_j}(\t_k-\t_j)\,, 
\ee
where 
\be
C_{V_j}=\langle\langle V_j(\t_j)\rangle\rangle = \exp\big(-\langle\langle \p^-_j(0)\p^+_j(0)\rangle\rangle  \big)\,,
\ee
\be
G_{V_kV_j}(\t_k-\t_j) = \exp\big(-\langle\langle \p_j(\t_j)\p_k(\t_k)\rangle\rangle  \big)\,.
\ee
The constants $C_{V_\mu}$ and the functions $G_{\mu\nu}\equiv G_{V_\mu V_\nu}$ are computed in appendix \ref{traces}. It is worth mentioning that some of the functions $G_{\mu\nu}$ are minimal two-particle form factors. In particular, $G_{(N-1)(1)}$ \eqref{GNm11} is the particle-antiparticle minimal form factor which determines 
the two-particle form factor of the current operator calculated in \cite{Cubero:2012xi}.

The integration contours in \eqref{Ffunc} are similar to the ones for the vacuum expectation values 
\be\la{Ffunccon}
\langle 0| Z_{M_n}(\t_{n})\cdots Z_{M_1}(\t_{1})|0 \rangle\,.\qquad
\ee
However, $G_{\mu|\nu}$-functions have more poles, and the rule for choosing an integration contour is modified as follows. In addition to the usual requirements, one also requires that an integration contour $C_{A}$ due to the operator $\chi=\int_C e^{i\p_A}$ is in the simply-connected region which contains all the poles of $g_{\mu|A}$-functions due to the vertex operators $V_\mu$ to the right of $\chi$, and all the poles of $g_{A|\nu}$-functions due to the vertex operators $V_\nu$ to the left of $\chi$ but no other poles of $G_{\mu|A}$ and $G_{A|\nu}$. 
For example, the integration contour $C_{k\dr}$ in $\chi^-_k$ runs from Re$\,\a_{k\dr}=-\infty$ to Re$\,\a_{k\dr}=+\infty$, and it lies above a pole of the $g_{\mu|{k\dr}}$-function due to the vertex operator $V_\mu$ to the right of $\chi^-_k$ and above all the poles of the $G_{\mu|{k\dr}}$-function which are below this pole of $g_{\mu|{k\dr}}$. However, $C_{k\dr}$ runs below all the poles of the $G_{\mu|{k\dr}}$-function which are above this pole of $g_{\mu|{k\dr}}$. If $g_{\mu|{k\dr}}$ has no pole then $C_{k\dr}$ just runs below all the poles of the $G_{\mu|{k\dr}}$-function. Similarly, the contour $C_{k\dr}$ runs
 below a pole of the $g_{{k\dr}|\nu}$-function due to the vertex operator $V_\nu$ to the left of $\chi^-_k$ and below all the poles of  $G_{{k\dr}|\nu}$ which are above this pole of $g_{{k\dr}|\nu}$ but above all the poles of  $G_{{k\dr}|\nu}$ which are below this pole of $g_{{k\dr}|\nu}$. If $g_{{k\dr}|\nu}$ has no pole then $C_{k\dr}$ just runs above all the poles of  $G_{{k\dr}|\nu}$, see appendix \ref{traces} for explicit examples.

%%%%%%%%%%%%%%%%
\section{Conclusion}

In this paper, a free field representation for the ZF algebra of the \sun2 PCF model was found.  
Similarly to the Fateev-Lashkevich representation \cite{FL} for the ZF algebra of the SS model, this representation allows one to construct form factors of $(N^2$-$1)$-parameter family of exponential fields of the \sun2 PCF model. The precise form of the exponential fields and their relation to the fields which appear in the Lagrangian of the model remain to be determined.

It is also useful to construct the operators $\L(\tilde O)$ which satisfy the equations \eqref{lamO}. They describe in particular the current and energy-momentum tensor operators. The construction of Lukyanov \cite{Lukyanov} of these operators does not seem to work for the PCF model.

 The approach developed should be applicable to any  two dimensional relativistic integrable model invariant  under a direct sum of two simple Lie or q-deformed algebras. It would be interesting to apply it to the other PCF models \cite{Ogievetsky:1987vv}.

An important problem which can be now addressed is to construct 
 a free field representation for the ZF algebra  of the \ads superstring sigma model in the light-cone gauge. The model is crossing invariant \cite{Janik} but it is not relativistic invariant which complicates its analytic properties.  The most difficult question is to find Green's function $g_{00}$. The results of this paper show that it would satisfy an extra crossing-type equation similar to \eqref{gogkr}, and hopefully one might be able to solve it.
 Since Green's functions $g_{\mu\nu}$ determine  functions $G_{\mu\nu}$ some of which play the role of minimal form factors, it is hoped that this approach may shed some light on the \ads form factors and their analytic properties.

%%%%%%%%%%%%%%%%
\section*{Acknowledgements}

I would like to  thank Tristan McLoughlin and Zoltan Bajnok for useful discussions, and the Galileo Galilei Institute for Theoretical Physics (GGI) for the hospitality and INFN for partial support during the completion of this work, within the program ``New Developments in AdS3/CFT2 Holography''

%%%%%%%%%%%%%%%%%%%%%%%%%
\appendix

\section{Green's functions derivation}\la{Greender}

Green's function $g_{00}(\t)$ is found from the requirement that it has a simple zero at $\t=0$ and no poles or zeroes for Im$(\t)<0$, and is given by \eqref{gPCF}
\be
g_{00}(\t)=g_{PCF}(\t)=\frac{ \Gamma
   \left(\frac{i \t}{2 \pi }-\frac{1}{N}+1\right) \Gamma
   \left(\frac{i \t}{2 \pi }+\frac{1}{N}\right)}{\Gamma \left(\frac{i
   \t}{2 \pi }\right) \Gamma
   \left(\frac{i \t}{2 \pi }+1\right)}\,,\quad f_{00}(t)=2{\sh{(N-1)\pi t\ov N}\ov \sh{\pi t}}\sh{\pi t \ov N}\,.
\ee

%%%%%%%%%%%%%%%%%%%%%%%%%%%%%%%%%%%%%%%
\subsection*{Left-left and right-right Green's functions}

Since up to an overall multiplier the operators $Z_{k\d1}$ (and $Z_{1\dk}$) form the ZF algebra of the \GN model, the S-matrices $S_{0|k\dr}(\t-\a_{k\dr})$, $S_{k\dr|l\dq}(\a_{k\dr}-\a_{l\dq})$ and $S_{0|\dk r}(\t-\a_{\dk r})$, $S_{\dk r|\dl q}(\a_{\dk r}-\a_{\dl q})$ are the same as the S-matrices $S_{0k}$ and $S_{kl}$ of the \GN model,  up to signs and shifts of $\a_{k\dr}$.
It is easy to determine the most general solution for the S-matrices from the ZF relations  \eqref{zzkdrldr} or \eqref{zzkdrkdq} and show that up to shifts of $\a_{k\dr}$ it is unique. 
 It is convenient to choose the following S-matrices
for  $S_{0|1\dr}$, $\dr=\d1,..., \dN$
\be\la{S01dr}
 S_{0|1\dr}(\t)=\frac{\theta}{\theta -\frac{2\pi i}{N} }\,,
\ee
 for $S_{k\dr|k\dq}$, $  k=1\,,..., N\,,\ \dr=\d1,..., \dN\,,$
 \be\la{Skdrkdr}
 S_{k\dr|k\dq}(\t)=\frac{\t-\frac{2\pi i}{N} }{\t+\frac{2 \pi i }{N} }\,,
\ee
and finally for $S_{k\dr|k+1\dq}$, $ k=1\,,..., N-1\,,\  \dr,\dq=\d1,\ldots, \dN\,,
$ 
 \be\la{Skdrk1dq}
 S_{k\dr|k+1,\dq}(\t)=\frac{\t}{\t-\frac{2 \pi i}{N} }\,,
 \ee
and the same expressions for $S_{\d1 r|\d1 q}$. All the other S-matrices are either related to the listed ones by unitarity or are equal to 1. This choice leads to a simple pole structure of the  
left-right Green's functions consistent with the discussion in subsection \ref{free1}.

The functions $g_{0|1\dr}$, $g_{1\dr|0}$ and $g_{k\dr|k+1,\dq}$, $g_{k+1,\dq|k,\dr}$, however, cannot be the same as $g_{k,k+1}=g_{k+1,k}$ functions for the \GN model because it would lead to the appearance of a pole at $\t_{12}=2\pi i/N$ 
in the products $Z_{1\d1}(\t_1)Z_{2\d1}(\t_2)$ and $Z_{1\d1}(\t_1)Z_{1\dt}(\t_2)$, and, say, $Z_{2\d1}(\t_1)Z_{3\d1}(\t_2)$ and $Z_{1\dt}(\t_1)Z_{1\dth}(\t_2)$ due to a pole in $g_{k,k+1}$ and $g_{k+1,k}$. Thus, one has to assume that only one of these functions has a pole. 

Since $\bA_{2\dr}^\dagger(\a)\bA^{1\dr}(\a)\bA_{1\d1}^\dagger(\t) \sim \de_{\d1\dr}\delta(\t-\a)$, one expects $ g_{0|1\d1}(\t)$ to have a pole at $\t=0$ while $g_{0|1\dr}(\t)$, $\dr\ge 2$ to be regular at $\t=0$. Then, the general ansatz for 
$g_{0|A}$ and $g_{A|0}$ functions is
\bal
 g_{0|1\d1}(\t)&={ \t-{2\pi i\ov N}\ov\t}\,h_{0|1\d1}(\t)\,, \quad g_{1\d1|0}(\t)=h_{0|1\d1}(-\t)\,, 
\\
g_{0|1\dr}(\t)&=h_{0|1\dr}(\t)\,,\quad g_{1\dr|0}(\t)=\frac{\theta }{\theta+\frac{2\pi i}{N} }\,h_{0|1\dr}(-\t)\,,\quad  
\dr\ge \dt\,,
\eal
\bal
 g_{0|\d1 1}(\t)&={ \t-{2\pi i\ov N}\ov\t}\,h_{0|\d1 1}(\t)\,, \quad g_{\d1 1|0}(\t)=h_{0|\d1 1}(-\t)\,, 
\\
g_{0|\d1 r}(\t)&=h_{0|\d1 r}(\t)\,,\quad g_{\d1 r|0}(\t)=\frac{\theta }{\theta+\frac{2\pi i}{N} }\,h_{0|\d1 r}(-\t)\,,\quad  
r\ge 2\,,\ 
\eal
where $h_{0|1\dr}(\t)$ cannot have poles and cannot have a zero at $\t=0$.  

Similarly, the general ansatz for 
the left-left and right-right Green's functions is
\be
g_{k\dr|k\dq}(\t)=h_{k\dr|k\dq}(\t)\,,\quad g_{k\dq|k\dr}(\t)={\theta+\frac{2\pi i}{N}\ov \theta -\frac{2\pi i}{N}}\,h_{k\dr|k\dq}(-\t)\,,\quad \dr< \dq\,,
\ee
\be
g_{k\dr|k\dr}(\t)={\theta\ov \theta -\frac{2\pi i}{N}}\,h_{k\dr|k\dr}(\t)\,,\quad h_{k\dr|k\dr}(-\t)=h_{k\dr|k\dr}(\t)\,,
\ee
\be
g_{\dk r|\dk q}(\t)=h_{\dk r|\dk q}(\t)\,,\quad g_{\dk q|\dk r}(\t)={\theta+\frac{2\pi i}{N}\ov \theta -\frac{2\pi i}{N}}\,h_{\dk r|\dk q}(-\t)\,,\quad r< q\,,
\ee
\be
g_{\dk r|\dk r}(\t)={\theta\ov \theta -\frac{2\pi i}{N}}\,h_{\dk r|\dk r}(\t)\,,\quad h_{\dk r|\dk r}(-\t)=h_{\dk r|\dk r}(\t)\,,
\ee
\bal
g_{k\dr|k+1,\dq}(\t)&={ \t-{2\pi i\ov N}\ov\t}\,h_{k\dr|k+1,\dq}(\t)\,, \quad g_{k+1,\dq|k\dr}(\t)=h_{k\dr|k+1,\dq}(-\t)\,, \quad\dr\ge \dq\,,
\\
g_{k\dr|k+1,\dq}(\t)&=h_{k\dr|k+1,\dq}(\t)\,, \quad g_{k+1,\dq|k,\dr}(\t)=\frac{\t}{\t+\frac{ 2\pi i}{N} }\,h_{k\dr|k+1,\dq}(-\t)\,, \quad\dr< \dq\,,
\eal
\bal
g_{\dk r|\dk+\d1,q}(\t)&={ \t-{2\pi i\ov N}\ov\t}\,h_{\dk r|\dk+\d1,q}(\t)\,, \quad g_{\dk+\d1,q|\dk r}(\t)=h_{\dk r|\dk+\d1,q}(-\t)\,, \quad r\ge q\,,
\\
g_{\dk r|\dk+\d1,q}(\t)&=h_{\dk r|\dk+\d1,q}(\t)\,, \quad g_{\dk+\d1,q|\dk r}(\t)=\frac{\t}{\t+\frac{ 2\pi i}{N} }\,h_{\dk r|\dk+\d1,q}(-\t)\,, \quad r< q\,.
\eal
The analytic properties of the $h$-functions will be discussed later.

All the other Green's functions which are not listed above have no poles. This follows from  the commutativity relations discussed in subsection \eqref{commutrel}.  They are set to 1.

%%%%%%%%%%%%%%%%%%%%%%%%%%%%%%%%%%%%%%%
\subsection*{Left-right  and right-left Green's functions}

The commutativity of left and right algebras charges implies that the corresponding Green's functions satisfy
\bal
&g_{k\dr|\dq n}(\a_l-\b_r)=g_{\dq n|k\dr}(\b_r-\a_l)\quad \Rightarrow\quad S_{k\dr|\dq n}(\a_l-\b_r)=1\,.
\eal 
Some of these functions  have no poles. 
The simplest choice is to take all nonsingular left-right functions to be equal to 1. 
 However,  according to the discussion in subsection \ref{free1}, the functions (grouped according to the relations below)
\bal
&(g_{k,\dn+\d1|\dn k}\,, g_{k \dn|\dn,k+1}) \,,\eal
have a pole at $\a_l=\b_r$ 
\bal
&g_{k,\dn+\d1|\dn k}(\a)={\a+{2\pi i\ov N}\ov \a}\,h_{k,\dn+\d1|\dn k}(\a)=g_{\dn k|k,\dn+\d1}(-\a)\,, \\
& g_{k\dn|\dn,k+1}(\a)={\a-{2\pi i\ov N}\ov \a}\,h_{k\dn|\dn,k+1}(\a)=g_{\dn,k+1|k\dn}(-\a)\,,
\eal
where $h$'s are functions which have no poles, and have no zero at $\a=0$.
It is easy to show that the commutativity  of left and right algebras charges then leads to the relations of the form
\bal\la{extrarel}
\p_{k,\dn+\d1}(\a)+\dphi_{\dn k}(\a)&=\p_{k\dn}(\a)+\dphi_{\dn,k+1}(\a)\,,\\
c_{k,\dn+1}(\a)c_{\dn k}(\a)r_{\dn k|k,\dn+1}&=c_{k\dn}(\a)c_{\dn, k+1}(\a)r_{k\dn|\dn,k+1}\,,\quad 1\le k\le N-1\,,\  \d1\le \dn\le \dN-\d1\,,
\eal
where $r_{k\dn|\dn,k+1}\equiv {\rm Res}(g_{k\dn|\dn,k+1}(\a))|_{\a=0}$, $r_{\dn k|k,\dn+1}\equiv {\rm Res}(g_{\dn k|k,\dn+1}(\a))|_{\a=0}$.

These relations  lead to a huge number of the following additional relations between various Green's functions
\bal\la{greenrel}
&g_{A|k,\dn+\d1}(\a)g_{A|\dn k}(\a) = g_{A|k \dn}(\a)g_{A|\dn,k+1}(\a)\,,\\
&g_{k,\dn+\d1|A}(\a)g_{\dn k|A}(\a) = g_{k \dn|A}(\a)g_{\dn,k+1|A}(\a)\,,
\eal
where $A$ is any of the indices of the fields. Taking the ratio of these equations, one gets the following relations between the S-matrices
\bal\la{Smatrel}
&S_{A|k,\dn+\d1}(\a)S_{A|\dn k}(\a) = S_{A|k \dn}(\a)S_{A|\dn,k+1}(\a)\,,
\eal
which are indeed satisfied. Thus, it is sufficient to consider just one set of these equations.

Now, taking $A=0$, one gets
\bal\la{greenrel2}
&g_{0|k,\dn+\d1}(\a)g_{0|\dn k}(\a) = g_{0|k\dn}(\a)g_{0|\dn,k+1}(\a)\,.
\eal
Nontrivial relations can appear only for $k=1$ or $\dn=1$. One gets for $k=1$ and $\dn\ge \dt$
\bal\la{greenrel3}
&g_{0|1,\dn+\d1}(\a)= g_{0|1\dn}(\a)\ \Rightarrow\ h_{0|1\dn}(\a)=h_{0|1\dt}(\a)\,,\quad \dn\ge\dt\,.
\eal
Similarly, for $k\ge 2$ and $\dn =\d1$
\bal\la{greenrel3b}
&g_{0|\d1 k}(\a) = g_{0|\d1,k+1}(\a)\ \Rightarrow\ h_{0|\d1 k}(\a)=h_{0|\d1 2}(\a)\,,\quad k\ge 2\,.
\eal
Finally,  for $k=1$ and $\dn=1$
\bal\la{greenrel3c}
&g_{0|1\dt}(\a)g_{0|\d1 1}(\a) = g_{0|1\d1}(\a)g_{0|\d1 2}(\a)\ \Rightarrow\  h_{0|1\dt}(\a)h_{0|\d1 1}(\a) = h_{0|1\d1}(\a)h_{0|\d1 2}(\a)\,,
\eal
and therefore there are three independent functions $h_{0|A}$.

Next, taking $A=m\dr$, one gets  the following relations between Green's functions
\bal\la{greenrelf}
&g_{m\dr|k,\dn+\d1}(\a)g_{m\dr|\dn k}(\a) = g_{m\dr|k\dn}(\a)g_{m\dr|\dn,k+1}(\a)\,,
\eal
There are several cases to be considered. 
\bee
\item $m\le k-2$ or $m\ge k+2$

The equations are  satisfied because all these Green's functions are equal to 1.
\item $m= k-1$: $\ g_{k-1,\dr|k,\dn+\d1}(\a)g_{k-1,\dr|\dn k}(\a) = g_{k-1,\dr|k\dn}(\a)\,.$ Then,
\bal
\hspace{-1cm}g_{k-1,\dn|k,\dn+\d1}(\a)g_{k-1,\dn|\dn k}(\a) = g_{k-1,\dn|k\dn}(\a)\ \Rightarrow\ h_{k-1,\dn|k,\dn+\d1}(\a)h_{k-1,\dn|\dn k}(\a) = h_{k-1,\dn|k\dn}(\a)\,,
\eal
\bal
g_{k-1,\dr|k,\dn+\d1}(\a)= g_{k-1,\dr|k\dn}(\a)\ \Rightarrow\ h_{k-1,\dr|k,\dn+\d1}(\a)= h_{k-1,\dr|k\dn}(\a) \quad {\rm if}\quad \dr\neq \dn\,.
\eal
\item $m= k+1$: $\ g_{k+1,\dr|k,\dn+\d1}(\a) = g_{k+1,\dr|k\dn}(\a)g_{k+1,\dr|\dn,k+1}(\a)\,.$ Then,
\bal
&g_{k+1,\dn+\d1|k,\dn+\d1}(\a) = g_{k+1,\dn+\d1|k\dn}(\a)g_{k+1,\dn+\d1|\dn,k+1}(\a)\ \Rightarrow\ \\
&h_{k,\dn+\d1|k+1,\dn+\d1}(\a) = h_{k\dn|k+1,\dn+\d1}(\a)h_{k+1,\dn+\d1|\dn,k+1}(-\a)\,.
\eal
\bal
g_{k+1,\dr|k,\dn+\d1}(\a) = g_{k+1,\dr|k\dn}(\a) \ \Rightarrow\  h_{k,\dn+\d1|k+1,\dr}(\a) = h_{k\dn|k+1,\dr}(\a)    \quad {\rm if}\quad  \dr\neq \dn+\d1\,.
\eal
Thus 
\bal
h_{k\dn|k+1,\dr}(\a)&=h_{k\dn|k+1,\d1}(\a)=h_{k\d1|k+1,\d1}(\a) \,, \quad \dn\ge \dr\,,\\
h_{k\dn|k+1,\dr}(\a)&=h_{k\d1|k+1,\dr}(\a)=h_{k\d1|k+1,\dt}(\a) \,, \quad \dn< \dr\,,\\
h_{k\dn|\dn,k+1}(\a)&=h_{k\d1|\d1,k+1}(\a)\,, \\
h_{k,\dn+1|\dn k}(\a)&=h_{k\dt|\d1 k}(\a) \,,\\
h_{k\d1|k+1,\d1}(\a)&=h_{k\d1|k+1,\dt}(\a)h_{k+1,\dt|\d1, k+1}(-\a)\,,\\
h_{k\d1|k+1,\d1}(\a)&=h_{k\d1|k+1,\dt}(\a)h_{k\d1|\d1,k+1}(\a)\,,\\
h_{k+1,\dt|\d1, k+1}(-\a)&=h_{k\d1|\d1,k+1}(\a)\,.
\eal

\item $m= k$: $\ g_{k\dr|k,\dn+\d1}(\a)g_{k\dr|\dn k}(\a) = g_{k\dr|k\dn}(\a)g_{k\dr|\dn,k+1}(\a)\,.$ Then,
\bal
g_{k\dn|k,\dn+\d1}(\a)= g_{k\dn|k\dn}(\a)g_{k\dn|\dn,k+1}(\a)\ \Rightarrow\ h_{k\dn|k,\dn+\d1}(\a)= h_{k\dn|k\dn}(\a)h_{k\dn|\dn,k+1}(\a)\,,
\eal
\bal
\hspace{-1cm}g_{k,\dn+\d1|k,\dn+\d1}(\a)g_{k,\dn+\d1|\dn k}(\a) = g_{k,\dn+\d1|k\dn}(\a)\ \Rightarrow\  h_{k,\dn+\d1|k,\dn+\d1}(\a)h_{k,\dn+\d1|\dn k}(\a) = h_{k,\dn+\d1|k\dn}(\a)\,.
\eal
\bal
g_{k\dr|k,\dn+\d1}(\a) = g_{k\dr|k\dn}(\a)\ \Rightarrow\ h_{k\dr|k,\dn+\d1}(\a) = h_{k\dr|k\dn}(\a) \quad {\rm if}\quad \dr\neq \dn \ {\rm or}\ \dr\neq \dn+\d1\,.
\eal
Thus,
\bal
h_{k\dn|k\dr}(\a)&=h_{k\d1|k\dt}(\a) \,, \quad \dn< \dr\,,\\
h_{k\dn|k\dn}(\a)&=h_{k\d1|k\d1}(\a) \,,\\
h_{k\d1|k\dt}(\a)&=h_{k\d1|k\d1}(\a)h_{k\d1|\d1, k+1}(\a) \,,\\
h_{k\d1|k\dt}(-\a)&=h_{k\d1|k\d1}(\a)h_{k\dt|\d1 k}(\a)\,,\\
h_{k\dt|\d1 k}(-\a)&=h_{k\d1|\d1,k+1}(\a)=h_{k+1,\dt|\d1, k+1}(-\a)=h_{k-1,\d1|\d1,k}(\a)=h_{1\d1|\d1 2}(\a)\,.
\eal
\eee
The functions $h_{\dn k|\dr q}$ satisfy similar relations. The simplest solution used in the paper is obviously $h_{A|B}=1$ for any $A,B$.

%%%%%%%%%%
\section{Constraints and elementary free fields} \la{solvconstr}
\subsection*{Solving the constraints} 
The constraints 
\bal
\p_{k\dr}(\t)+\dphi_{\dr,k+1}(\t)=\dphi_{\dr k}(\t)+\p_{k,\dr+\d1}(\t)
\eal 
between the fields can be easily solved. In terms of the new fields
\bal
\vp_{k\d1}=\p_{k\d1}\,,\ \vp_{k\dt}=\p_{k\dt}-\p_{k\d1}\,,\ \vp_{k\dth}=\p_{k\dth}-\p_{k\dt}\,,\ldots\,, \ \vp_{k\dN}=\p_{k\dN}-\p_{k,\dN-\d1}\,,
\eal
\bal
\vp_{k\dr}=\p_{k\dr}-\p_{k,\dr-\d1}\,,\quad \p_{k\dot 0}=0\,,\quad k=1,\ldots,N-1\,,\quad \dr=\d1,\ldots, \dN\,,
\eal
\bal
\p_{k\d1}=\vp_{k\d1}\,,\ \p_{k\dt}=\vp_{k\dt}+\vp_{k\d1}\,,\ \p_{k\dth}=\vp_{k\dth}+\vp_{k\dt}+\vp_{k\d1}\,,\ldots\,, 
\eal
\bal
\p_{k\dr}=\sum_{\dq=\d1}^\dr\vp_{k\dq}\,,\quad k=1,\ldots,N-1\,,\quad \dr=\d1,\ldots, \dN\,,
\eal
\bal
\dot\vp_{\dk 1}=\dot\p_{\dk 1}\,,\ \dot\vp_{\dk 2}=\dot\p_{\dk 2}-\dot\p_{\dk 1}\,,\ \dot\vp_{\dk 3}=\dot\p_{\dk 3}-\dot\p_{\dk 2}\,,\ldots\,, \ \dot\vp_{\dk N}=\dot\p_{\dk N}-\dot\p_{\dk,N-1}\,,
\eal
\bal
\dot\vp_{\dk r}=\dot\p_{\dk r}-\dot\p_{\dk,r-1}\,,\quad \dot\p_{\dk 0}=0\,,\quad \dk=\d1,\ldots,\dN-\d1\,,\quad r=1,\ldots, N\,,
\eal
\bal
\dot\p_{\dk 1}=\dot\vp_{\dk 1}\,,\ \dot\p_{\dk 2}=\dot\vp_{\dk 2}+\dot\vp_{\dk 1}\,,\ \dot\p_{\dk 3}=\dot\vp_{\dk 3}+\dot\vp_{\dk 2}+\dot\vp_{\dk 1}\,,\ldots\,, 
\eal
\bal
\dot\p_{\dk r}=\sum_{q=1}^r\dot\vp_{\dk q}\,,\quad \dk=\d1,\ldots,\dN-\d1\,,\quad r=1,\ldots, N\,,
\eal
the constraints take the form
\bal
\vp_{k,\dr+\d1} = \dot\vp_{\dr,k+1} \equiv \psi_{kr}\,,\quad k=1,\ldots,N-1\,,\quad r=1,\ldots, N-1\,,
\eal
and the solution is written in terms of $\p_{k\d1}$, $\dot\p_{\dk 1}$, and $\psi_{kr}$
\bal
\p_{k\dr}=\p_{k\d1}+\sum_{q=1}^{r-1}\psi_{kq}=\sum_{q=0}^{r-1}\psi_{kq}\,,\quad  \dot\p_{\dk r}=\dot\p_{\dk 1}+\sum_{q=1}^{r-1}\psi_{qk}=\sum_{q=0}^{r-1}\psi_{qk}\,,
\eal
where
\bal
\psi_{k0}\equiv \p_{k\d1}\,,\quad \psi_{0k}\equiv \dot\p_{\dk 1}\,.
\eal
Green's functions of $\psi_{kq}$'s are given by
\bal
g^{\psi\psi}_{kr|nq}(\t) = {g_{k,\dr+\d1|n,\dq+\d1}(\t)\,g_{k\dr|n\dq}(\t)\ov g_{k,\dr+\d1|n\dq}(\t)\,g_{k\dr|n,\dq+\d1}(\t)}\,,\quad k\,, n \ge 1\,,
\eal
\bal
g^{\psi\psi}_{0r|nq}(\t) =g^{\phi\psi}_{\dr 1|nq}(\t) = {g_{\dr 1|n,\dq+\d1}(\t)\ov g_{\dr 1|n\dq}(\t)}\,,\quad  n \ge 1\,,
\eal
\bal
g^{\psi\psi}_{kr|0q}(\t) =g^{\psi\phi}_{kr|\dq 1}(\t) = {g_{k,\dr+\d1|\dq 1}(\t)\ov g_{k\dr|\dq 1}(\t)}\,,\quad k\ge 1\,,
\eal
\bal
g^{\psi\psi}_{0r|0q}(\t) =g_{\dr 1|\dq 1}(\t) \,.
\eal
The functions different from 1 are
\bal
g^{\psi\psi}_{kr|k+1,r+1}(\t) = {g_{k,\dr+\d1|k+1,\dr+\dt}(\t)\,g_{k\dr|k+1,\dr+\d1}(\t)\ov g_{k,\dr+\d1|k+1,\dr+\d1}(\t)\,g_{k\dr|k+1,\dr+\dt}(\t)}={\theta\ov \theta -\frac{2\pi i}{N}}\,,\quad k \ge 1\,,\ \dr\ge \dot 0\,,
\eal
\bal
g^{\psi\psi}_{k+1,r+1|kr}(\t) = {g_{k+1,\dr+\dt|k,\dr+\d1}(\t)\,g_{k+1,\dr+\d1|k\dr}(\t)\ov g_{k+1,\dr+\dt|k,\dr}(\t)\,g_{k+1,\dr+1|k,\dr+1}(\t)}={\theta\ov \theta +\frac{2\pi i}{N}}\,,\quad k \ge 1\,,\ \dr\ge \dot 0\,,
\eal
\bal
g^{\psi\psi}_{kr|k+1,r}(\t) = {g_{k,\dr+\d1|k+1,\dr+\d1}(\t)\,g_{k\dr|k+1,\dr}(\t)\ov g_{k,\dr+\d1|k+1,\dr}(\t)\,g_{k\dr|k+1,\dr+\d1}(\t)}={\theta-\frac{2\pi i}{N}\ov \theta }\,,\quad k \ge 1\,,\ \dr\ge \dot 0\,,
\eal
\bal
g^{\psi\psi}_{k+1,r|kr}(\t) = {g_{k+1,\dr+\d1|k,\dr+\d1}(\t)\,g_{k+1,\dr|k\dr}(\t)\ov g_{k+1,\dr+\d1|k\dr}(\t)\,g_{k+1,\dr|k,\dr+\d1}(\t)}={\theta+\frac{2\pi i}{N}\ov \theta }\,,\quad k \ge 1\,,\ \dr\ge \dot 1\,,
\eal
\bal
g^{\psi\psi}_{kr|k,r+1}(\t) = {g_{k,\dr+\d1|k,\dr+\dt}(\t)\,g_{k\dr|k,\dr+1}(\t)\ov g_{k,\dr+\d1|k,\dr+1}(\t)\,g_{k\dr|k,\dr+\dt}(\t)}={\theta-\frac{2\pi i}{N}\ov \theta }\,,\quad k \ge 1\,,\ \dr\ge \dot 0\,,
\eal
\bal
g^{\psi\psi}_{k,r+1|k\dr}(\t) = {g_{k,\dr+\dt|k,\dr}(\t)\,g_{k,\dr+\d1|k\dr}(\t)\ov g_{k,\dr+\dt|k\dr}(\t)\,g_{k,\dr+\d1|k,\dr+\d1}(\t)}={\theta+\frac{2\pi i}{N}\ov \theta }\,,\quad k \ge 1\,,\ \dr\ge \dot 0\,,
\eal
\bal
g^{\psi\psi}_{kr|kr}(\t) = {g_{k,\dr+\d1|k,\dr+\d1}(\t)\,g_{k\dr|k\dr}(\t)\ov g_{k,\dr+\d1|k\dr}(\t)\,g_{k\dr|k,\dr+\d1}(\t)}={\theta^2\ov \theta^2 +\frac{4\pi^2}{N^2} }\,,\quad k \ge 1\,,\ \dr\ge \dot 1\,,
\eal
\bal
g^{\psi\psi}_{k0|k0}(\t) = g_{k\d1|k\d1}(\t)={\theta\ov \theta-\frac{2\pi i}{N} }\,,\quad k \ge 1\,,
\eal
\bal
g^{\psi\psi}_{0r|1r}(\t) =g^{\phi\psi}_{\dr 1|1r}(\t) = {g_{\dr 1|1,\dr+\d1}(\t)\ov g_{\dr 1|1\dr}(\t)}={\theta-\frac{2\pi i}{N}\ov \theta }\,,\quad  \dr \ge \d1\,,
\eal
\bal
g^{\psi\psi}_{1r|0r}(\t) =g^{\psi\phi}_{1r|\dr 1}(\t) = {g_{1,\dr+\d1|\dr 1}(\t)\ov g_{1\dr|\dr 1}(\t)}={\theta+\frac{2\pi i}{N}\ov \theta }\,,\quad \dr\ge \d1\,,
\eal
\bal
g^{\psi\psi}_{0r|1,r+1}(\t) =g^{\phi\psi}_{\dr 1|1,r+1}(\t) = {g_{\dr 1|1,\dr+\dt}(\t)\ov g_{\dr 1|1,\dr+\d1}(\t)}={\theta\ov \theta -\frac{2\pi i}{N}}\,,\quad  \dr \ge \d1\,,
\eal
\bal
g^{\psi\psi}_{1,r+1|0r}(\t) =g^{\psi\phi}_{1,r+1|\dr 1}(\t) = {g_{1,\dr+\dt|\dr 1}(\t)\ov g_{1,\dr+\d1|\dr 1}(\t)}={\theta\ov \theta +\frac{2\pi i}{N}}\,,\quad \dr\ge \d1\,,
\eal
\bal
g^{\psi\psi}_{0r|0,r+1}(\t) =g_{\dr 1|\dr+\d1, 1}(\t) ={\theta-\frac{2\pi i}{N}\ov \theta } \,,\quad \dr\ge \dot 1\,,
\eal
\bal
g^{\psi\psi}_{0r|0r}(\t) =g_{\dr 1|\dr 1}(\t) ={\theta\ov \theta -\frac{2\pi i}{N}}\,,\quad \dr\ge \dot 1 \,.
\eal
It is interesting that the only Green's functions with S-matrices different from 1 are 
$g^{\psi\psi}_{0r|0,r+1}\,,\, g^{\psi\psi}_{0r|0r}\,,\, g^{\psi\psi}_{k0|k+1,0}\,,\, g^{\psi\psi}_{k0|k0}$.

%%%%%%%%%%%%%%%%%%%%%%%
\subsection*{Elementary free fields}

It seems that the best way to handle the free fields is to introduce $N^2-1$ elementary fields $\xi_{kr}(\t)$, $k,r=0,1,\ldots N-1$,  $\xi_{00}(\t)=0$, which satisfy the simplest commutation relations
\bal\la{elemff}
\xi_{kr}(\t)=\int_{-\infty}^\infty {dt\ov i t} a_{kr}(t)e^{i\t t} =\xi_{kr}^-(\t)+\xi_{kr}^+(\t)\,,\quad [a_{kr}(t),a_{nq}(t')]=t\de_{kr,nq}\de(t+t') \,,
\eal
 \bal
\bra\xi_{nq}(\t_1)\xi_{kr}(\t_2)\ket=-\log g_{kr|nq}(\t_{21}) = -\de_{kr,nq}\log i\,e^\g\t_{21}\,,\quad g_{kr|kr}(\t) =i\,e^\g \t\,,
\eal
where $\g$ is  Euler's constant.

There are infinitely many  different ways to represent $\psi_{kr}$ in terms of the elementary fields. By using the ansatz
\bal
\psi_{kr}^+(\t) &= \xi_{kr}^+(\t)\,,\\
\psi_{kr}^-(\t) &= \sum_A \Big(d_{kr|A}\,(\xi_A^-(\t-\un_2)-\xi_A^-(\t)) +f_{kr|A}\,(\xi_A^-(\t+\un_2)-\xi_A^-(\t))  \Big)\,,
\eal
where $\xi_A$ are the $N^2-1$ elementary fields, one finds the following representation
\bal
\psi_{kr}^-(\t) &= 2\xi_{kr}^-(\t)-\xi_{kr}^-(\t-\un_2) -\xi_{kr}^-(\t+\un_2) \\
&
+ \xi_{k-1,r-1}^-(\t)-\xi_{k-1,r-1}^-(\t+\un_2)+\xi_{k+1,r+1}^-(\t)-\xi_{k+1,r+1}^-(\t-\un_2) \\
&-\xi_{k+1,r}^-(\t)+
\xi_{k+1,r}^-(\t-\un_2)-\xi_{k-1,r}^-(\t)+
\xi_{k-1,r}^-(\t+\un_2)\\
&-\xi_{k,r+1}^-(\t)+
\xi_{k,r+1}^-(\t-\un_2) -\xi_{k,r-1}^-(\t)+
\xi_{k,r-1}^-(\t+\un_2)\,,\\
\psi_{kr}^+(\t) &= \xi_{kr}^+(\t)\,, \quad k \ge 2\,,\ r\ge  1 \,,
\eal
\bal
\psi_{k0}^-(\t) &=\xi_{k0}^-(\t) -\xi_{k0}^-(\t+\un_2) -\xi_{k1}^-(\t)+
\xi_{k1}^-(\t-\un_2)  \\
&+\xi_{k+1,1}^-(\t)
-\xi_{k+1,1}^-(\t-\un_2)-\xi_{k-1,0}^-(\t) +
\xi_{k-1,0}^-(\t+\un_2)\,,\\
\psi_{k0}^+(\t) &= \xi_{k0}^+(\t)\,, \quad k \ge 2\,,\hspace{7.8cm}
\eal
\bal
\psi_{10}^-(\t) &=\xi_{10}^-(\t) -\xi_{10}^-(\t+\un_2) -\xi_{11}^-(\t)+
\xi_{11}^-(\t-\un_2)  \\
&+\xi_{21}^-(\t)
-\xi_{21}^-(\t-\un_2)\,,\\
\psi_{10}^+(\t) &= \xi_{10}^+(\t)\,, \hspace{9.5cm}
\eal
\bal
\psi_{1r}^-(\t) &= 2\xi_{1r}^-(\t)-\xi_{1r}^-(\t-\un_2) -\xi_{1r}^-(\t+\un_2)  \\
&+\xi_{0,r-1}^-(\t)
 -\xi_{0,r-1}^-(\t+\un_2)+\xi_{2,r+1}^-(\t)-\xi_{2,r+1}^-(\t-\un_2) \\
&-\xi_{2r}^-(\t)+
\xi_{2r}^-(\t-\un_2)-\xi_{0r}^-(\t)+
\xi_{0r}^-(\t+\un_2)\\
&-\xi_{1,r+1}^-(\t)+
\xi_{1,r+1}^-(\t-\un_2) -\xi_{1,r-1}^-(\t)+
\xi_{1,r-1}^-(\t+\un_2) \,,\\
\psi_{1r}^+(\t) &= \xi_{1r}^+(\t)\,, \quad \ r\ge  2\,,\hspace{8cm}
\eal
\bal
\psi_{11}^-(\t) &= 2\xi_{11}^-(\t)-\xi_{11}^-(\t-\un_2) -\xi_{11}^-(\t+\un_2) +\xi_{22}^-(\t)-\xi_{22}^-(\t-\un_2) \\
&-\xi_{21}^-(\t)+
\xi_{21}^-(\t-\un_2)-\xi_{01}^-(\t)+
\xi_{01}^-(\t+\un_2)\\
&-\xi_{12}^-(\t)+
\xi_{12}^-(\t-\un_2) -\xi_{10}^-(\t)+
\xi_{10}^-(\t+\un_2) \,,\\
\psi_{11}^+(\t) &= \xi_{11}^+(\t)\,,\hspace{9.6cm}
\eal
\bal
\psi_{0r}^-(\t) &= \xi_{0r}^-(\t)-\xi_{0r}^-(\t+\un_2) +\xi_{1,r+1}^-(\t)
-\xi_{1,r+1}^-(\t-\un_2) \\
&-\xi_{1r}^-(\t)+
\xi_{1r}^-(\t-\un_2)-\xi_{0,r-1}^-(\t)+
\xi_{0,r-1}^-(\t+\un_2)
\,,\\
\psi_{0r}^+(\t) &= \xi_{0r}^+(\t)\,, \quad \ r\ge  2\,,\hspace{8cm}
\eal
\bal
\psi_{01}^-(\t) &= \xi_{01}^-(\t)-\xi_{01}^-(\t+\un_2) +\xi_{12}^-(\t)
-\xi_{12}^-(\t-\un_2) \\
&-\xi_{11}^-(\t)+
\xi_{11}^-(\t-\un_2)
\,,\\
\psi_{01}^+(\t) &= \xi_{01}^+(\t)\,.\hspace{9.5cm}
\eal
These formulae are used to express $\p_{k\dr}$ and $\p_{\dk r}$ in terms of the elementary fields
\bal\la{phidrxi}
\p^-_{k\dr}(\t)&=\xi^-_{k,r-1}(\t)-\xi^-_{k,r-1}(\t+\un_2) -\xi^-_{k-1,r-1}(\t)+\xi^-_{k-1,r-1}(\t+\un_2)\\
&+
\xi^-_{k+1,r}(\t)-\xi^-_{k+1,r}(\t-\un_2) -\xi^-_{kr}(\t)+\xi^-_{kr}(\t-\un_2)\,,\\
%\eal
%\bal
\dot\p^-_{\dk r}(\t)&=\xi^-_{r-1,k}(\t)-\xi^-_{r-1,k}(\t+\un_2) -\xi^-_{r-1,k-1}(\t)+\xi^-_{r-1,k-1}(\t+\un_2)\\
&-
\xi^-_{rk}(\t)+\xi^-_{rk}(\t-\un_2) +\xi^-_{r,k+1}(\t)-\xi^-_{r,k+1}(\t-\un_2)\,,\\
%\eal
%\bal
\p^+_{k\dr}(\t)&=\sum_{q=0}^{r-1}\xi^+_{kq}(\t)\,,\quad  \dot\p^+_{\dk r}(\t)=\sum_{q=0}^{r-1}\xi^+_{qk}(\t)\,,
\eal
and therefore (summing over $m,n$)
\bal\la{phidrxi2}
\p^-_{k\dr}(\t)&=\int_{0}^\infty\, {dt\ov it}\,\Phi_{k\dr,mn}^-(t)a_{mn}(t)\, e^{i\t t}\,,\\
 \Phi_{k\dr, mn}^-(t)&= (1-e^{-{2\pi\ov N}t})(\de_{km}\de_{r-1,n}-\de_{k-1,m}\de_{r-1,n})+ (1-e^{{2\pi\ov N}t})(\de_{k+1,m}\de_{rn}-\de_{km}\de_{rn})\,,\\
%\eal
%\bal
\dot\p^-_{\dk r}(\t)&=\int_{0}^\infty\, {dt\ov it}\,\Phi_{\dk r,mn}^-(t)a_{mn}(t)\, e^{i\t t}\,,\quad \Phi_{\dk r, mn}^-(t)=  \Phi_{k\dr, nm}^-(t)\,,
\eal
\bal
\p^+_{k\dr}(\t)&=\int^{0}_{-\infty}\, {dt\ov it}\,\Phi_{k\dr,mn}^+(t)a_{mn}(t)\, e^{i\t t}\,,  \quad \Phi_{k\dr, mn}^+(t)=\de_{km}(1-u(n-r))\,,\\
\p^+_{\dk r}(\t)&=\int^{0}_{-\infty}\, {dt\ov it}\,\Phi_{\dk r,mn}^+(t)a_{mn}(t)\, e^{i\t t}\,,  \quad \Phi_{\dk r, mn}^+(t)=\de_{kn}(1-u(m-r))\,,
\eal
where $u(x)$ is the unit step function
\bal
u(x) = \left\{ \begin{array}{ccc}
\ 0\ &\ {\rm for}  &\ x<0  \\
\ 1\ &\   {\rm for}  &\ x\ge 0  \end{array}
 \right.\,.
\eal

%%%%%%%%%%%%%%%%%%%%%%%
\subsection*{$\p_{0}$ and $\p_{(r)}$ in terms of elementary free fields}

Then one gets $\tilde\p_{(r)}$ which appears in the rank-$r$ heighest weight bound state operator 
\bal
\p_{(r)} (\t)&=\sum_{k=-{r-1\ov 2}}^{{r-1\ov 2}}\p_0(\t+\un_{2k})+\tilde\p_{(r)}(\t) \,,\\
\tilde\p_{(r)}(\t) &=\sum_{n=1}^{r-1}\sum_{k=-{r-1\ov 2}}^{{r-1\ov 2}-n}\big(\p_{n\dn}(\t+\un_{2k})+\p_{\dn, n+1}(\t+\un_{2k})\big)\,,\\
\tilde\p^-_{(r)} (\t)&=\xi_{11}^-\left(\t-\un_{r+1} \right)-\xi_{11}^-\left(\t+\un_{r-1}\right)+\xi_{rr}^-\left(\t-\un_{r-1}\right)-\xi_{rr}^-\left(\t-\un_{r+1}\right)\\
\tilde\p^-_{(r)}(\t)&=\int_{0}^\infty\, {dt\ov it}\,\tilde\Phi_{(r),mn}^-(t)a_{mn}(t)\, e^{i\t t}\,,\\
\tilde \Phi_{(r), mn}^-(t)&= e^{{\pi\ov N}t}(e^{{\pi r\ov N}t}-e^{-{\pi r\ov N}t})\de_{1m}\de_{1n}+e^{{\pi r\ov N}t}(e^{-{\pi\ov N}t}-e^{{\pi\ov N}t})\de_{rm}\de_{rn}
\,.
\eal
Taking into account that
\bal
\sum_{k=-{r-1\ov 2}}^{{r-1\ov 2}-n}\p_{\mu}(\t+\un_{2k})=\int_{-\infty}^\infty\, {dt\ov it}\,e^{\frac{\pi  n}{N}t}  {\sinh\pi t{r-n\ov N}\ov \sinh {\pi t\ov N}}\,\Phi_{\mu A}(t)a_A(t)\, e^{i\t t}\,,
\eal
one gets
\bal
\tilde\p^+_{(r)}(\t)&=\int^{0}_{-\infty}\, {dt\ov it}\,\tilde\Phi_{(r),mn}^+(t)a_{mn}(t)\, e^{i\t t}\,,  \\
\tilde \Phi_{(r), mn}^+(t)&=e^{\frac{\pi  m}{N}t}  {\sinh\pi t{r-m\ov N}\ov \sinh {\pi t\ov N}}u(r-1-m)u(r-1-n)u(m-1-n)\\
 &+e^{\frac{\pi  n}{N}t}  {\sinh\pi t{r-n\ov N}\ov \sinh {\pi t\ov N}}u(r-1-m)u(r-1-n)u(n-m)\,,
\eal
and therefore
\bal
\tilde\p^+_{(r)}(\t)&=\int^{0}_{-\infty}\, {dt\ov it}\,\sum_{m,n=0}^{r-1}f_{mn}^{(r)}(t)a_{mn}(t)\, e^{i\t t}\,,  \\
 f_{mn}^{(r)}(t)&= \left\{ \begin{array}{ccc}
\ e^{\frac{\pi  m}{N}t}  {\sinh\pi t{r-m\ov N}\ov \sinh {\pi t\ov N}}\ &\ {\rm for}  &\ n<m<r  \\
\ e^{\frac{\pi  n}{N}t}  {\sinh\pi t{r-n\ov N}\ov \sinh {\pi t\ov N}}\ &\   {\rm for}  &\ m\le n<r  \end{array}
 \right.\,,
\eal
\bal
\p_{(r)} (\t)&=\int_{-\infty}^\infty\, {dt\ov it}\,{\sinh\pi t{r\ov N}\ov \sinh {\pi t\ov N}}\,A_{0}(t)\, e^{i\t t}+\tilde\p_{(r)}(\t)\,.
\eal
Then the relation $V_{(N)} =1$ becomes $\p_{(N)} =0$, and allows one to express $\p_0^\pm$ in terms of $\xi_{kr}^\pm$. Taking into account that $\xi_{NN}=0$, and that
\bal
\sum_{k=-{N-1\ov 2}}^{{N-1\ov 2}}\p_0(\t+\un_{2k})=\int_{-\infty}^\infty\, {dt\ov it}\,{\sinh \pi t\ov \sinh {\pi t\ov N}}A_{0}(t)\, e^{i\t t}=\int_{-\infty}^\infty\, {dt\ov it}\,{\sinh \pi t\ov \sinh {\pi t\ov N}}\,\Phi_{0,mn}(t)a_{mn}(t)\, e^{i\t t}\,,
\eal
one gets
\bal
\Phi^-_{0,mn}(t)= -2e^{{\pi\ov N}t}\sinh {\pi t\ov N}\,\de_{1m}\de_{1n}\ \Rightarrow\ A_0^-(t) =  -2e^{{\pi\ov N}t}\sinh {\pi t\ov N}a_{11}(t)\,,
\eal
\bal
\Phi^+_{0,mn}(t)= -{\sinh {\pi t\ov N}\ov\sinh \pi t}f_{mn}^{(N)}(t) =  \left\{ \begin{array}{ccc}
-\ e^{\frac{\pi  m}{N}t}  {\sinh\pi t{N-m\ov N}\ov \sinh {\pi t}}\ &\ {\rm for}  &\ n<m  \\
-\ e^{\frac{\pi  n}{N}t}  {\sinh\pi t{N-n\ov N}\ov \sinh {\pi t}}\ &\   {\rm for}  &\ n\ge m  \end{array}
 \right. \,.
\eal
Thus, 
\bal
\big[ A_0(t),A_0(t')\big]&=\Phi^-_{0,mn}(t)\Phi^+_{0,mn}(-t)\,t\de(t+t')= t f_{00}(t)\de(t+t')\,,\\
f_{00}(t) &=\Phi^-_{0,mn}(t)\Phi^+_{0,mn}(-t)= 2{\sinh {1\ov N}\pi t\,\sinh {N-1\ov N}\pi t\ov\sinh \pi t}\,,\quad t>0\,,
\eal
as required.
Moreover, by using the formulae for $A_0$, one gets
\bal
\p_{(r)}^- (\t)&=\int_{0}^\infty\, {dt\ov it}\,\Phi_{(r),mn}^-(t)a_{mn}(t)\, e^{i\t t}\,,\\
\Phi_{(r), mn}^-(t)&=-2e^{{\pi r\ov N}t}\sinh{{\pi\ov N}t}\,\de_{rm}\de_{rn}\ \Rightarrow\ A_{(r)}^-(t) =  -2e^{{\pi r\ov N}t}\sinh {\pi t\ov N}a_{rr}(t)\,,
\eal
\bal
\p_{(r)}^+ (\t)&=\int^{0}_{-\infty}\, {dt\ov it}\,\Phi_{(r),mn}^+(t)a_{mn}(t)\, e^{i\t t}\,,\quad
\Phi_{(r), mn}^+(t)=-{\sinh{{\pi r\ov N}t}\ov\sinh{{\pi}t}}\,f_{mn}^{(N)}(t)+\tilde\Phi_{(r), mn}^+(t)\,.
\eal
Notice that $\p_{(1)}=\p_0$, and these formulae are consistent with the equality.
In particular, one gets 
\bal
\big[ A_{(r)}(t),A_{(r)}(t')\big]&=\Phi^-_{(r),mn}(t)\Phi^+_{(r),mn}(-t)\,t\de(t+t')= t f_{(r)(r)}(t)\de(t+t')\,,\quad t>0\,,\\
f_{(r)(r)}(t) &=\Phi^-_{(r),mn}(t)\Phi^+_{(r),mn}(-t)= 2{\sinh { r\ov N}\pi t\,\sinh {N-r\ov N}\pi t\ov\sinh \pi t}\,,\quad t>0\,,
\eal
and in general
\bal
\big[ A_{(r)}(t),A_{(q)}(t')\big]&=\Phi^-_{(r),mn}(t)\Phi^+_{(q),mn}(-t)\,t\de(t+t')= t f_{(r)(q)}(t)\de(t+t')\,,\quad t>0\,,\\
f_{(r)(q)}(t) & =\Phi^-_{(r),rr}(t)\Phi^+_{(q),rr}(-t)
=  \left\{ \begin{array}{ccc}
2{\sinh { q\ov N}\pi t\,\sinh {N-r\ov N}\pi t\ov\sinh \pi t}\ &\ {\rm for}  &\ r\ge q  \\
2{\sinh { r\ov N}\pi t\,\sinh {N-q\ov N}\pi t\ov\sinh \pi t}\ &\   {\rm for}  &\ r<q  \end{array}
 \right.
\,,\quad t>0\,,
\eal
where it is used that 
\bal
2{\sinh { q\ov N}\pi t\,\sinh {N-r\ov N}\pi t\ov\sinh \pi t}-2\sinh\pi t{q-r\ov N}=2{\sinh { r\ov N}\pi t\,\sinh {N-q\ov N}\pi t\ov\sinh \pi t}\,.
\eal
Then, one gets
\bal
g_{(r)(q)}(\t )=g_{(q)(r)}(\t )\,,\quad g_{(r)(q)}(\t )=\frac{\Gamma \left(\frac{i \t}{2 \pi }-\frac{r+q}{ 2N}+1\right) \Gamma \left(\frac{i
   \t}{2 \pi }+\frac{r+q}{ 2N}\right)}{\Gamma \left(\frac{i \t}{2 \pi }-\frac{r-q}{
   2N}+1\right) \Gamma \left(\frac{i \t}{2\pi
   }+\frac{r-q}{2N}\right)}\,,\quad 
   r\ge q\,.
\eal

%%%%%%%%%%%%%%%%
 \section{ZF operators $Z_{2\d1}$, $Z_{1\dt}$, $Z_{2\dt}$} \la{Z12op}
 
 It is  easy to get $Z_{2\d1}$ and $Z_{1\dt}$
\bal\la{z2d1}
Z_{2\d1}(\t)&=Z_{2\d1}^{(1)}(\t)+Z_{2\d1}^{(2)}(\t) \\
&= \r(\t)\int_{\t^{--}/C_{1\dr}} \,c_{1\dr}\,g^a_{0|1\dr}\,e^{i\p_0+i\p_{1\dr}} + \r(\t)c_{1\dr}(\t)\,R_{0|1\dr}\,e^{i\p_0+i\p_{1\dr}(\t)}\,,
\eal
\bal\la{z1d2}
Z_{1\dt}(\t)&=Z_{1\dt}^{(1)}(\t)+Z_{1\dt}^{(2)}(\t)\\
&=  \r(\t)\int_{\t^{--}/C_{\d1 r}}\,c_{\d1 r}\,g^a_{0|\d1 r}\,e^{i\p_0+i\p_{\d1 r}} +  \r(\t)c_{\d1 r}(\t)R_{0|\d1 r}\,e^{i\p_0(\t)+i\dphi_{\d1 r}(\t)}\,.
\eal
 Taking into account that $R_{0|1 \dr}=\de_{\dr\d1}R_{0|1\d1}$, and $R_{0|\d1 r}=\de_{r1}R_{0|\d1 1}$, one obtains \eqref{Z2d1} and \eqref{Z1d2}.

Before shifting the contours the operator $Z_{2\dt}$ is given by 
\bal\la{z2d20}
Z_{2\dt}(\t)= \dchi_\d1^-Z_{2\d1}(\t) -  Z_{2\d1}(\t)\dchi_\d1^-  \,.
\eal
The second term does not require any shifting of the contours, and is given by 
\bal\la{z2d2t2}
-Z_{2\d1} \dchi_\d1^- &= -  \r(\t)\int_{\t^{--}/C_{1\dr}/C_{\d1 q}}  \, c_{1\dr}\, c_{\d1 q}\,g^a_{0|1\dr}g_{\d1 q|0}\, g_{\d1 q|1\dr}\,e^{i\p_0+i\phi_{1\dr}+i\dphi_{\d1 q}}
\\
&-  \r(\t) \int_{\t^{--}/C_{\d1 q}}  \, c_{\d1 q}\,c_{1\dr}(\t_{1\dr})\,R_{0|1\dr}\,g_{\d1 q|0}\, g_{\d1 q|1\dr}(\a_{\d1 q}-\t)\,e^{i\p_0+i\p_{1\dr}(\t)+i\dphi_{\d1 q}}\,.
 \eal
The first term is given by 
\bal\la{z2d21}
 \dchi_\d1^-Z_{2\d1} &= \dchi_\d1^- Z_{2\d1}^{(1)}(\t)+ \dchi_\d1^- Z_{2\d1}^{(2)}(\t)
 \\
 &=  \r(\t) \int_{C_{\d1 q}/\t/\t^{--}/C_{1\dr}} \,c_{1\dr} \,c_{\d1 q}\,g^a_{0|1\dr}\,g_{0|\d1 q}\, g_{1\dr|\d1 q}\,e^{i\p_0+i\phi_{1\dr}+i\dphi_{\d1 q}}
 \\
 &+  \r(\t)\int_{C_{\d1 q}/\t} \,c_{1\dr}(\t) \,c_{\d1 q}\,R_{0|1\dr}\,g_{0|\d1 q}(\t- \a_{\d1 q})\, g_{1\dr|\d1 q}(\t- \a_{\d1 q})\,e^{i\p_0+i\phi_{1\dr}(\t)+i\dphi_{\d1 q}}\,.
 \eal
 Shifting $C_{\d1 q}$ down, one gets
\bal\la{z2d23}
 \dchi_\d1^- Z_{2\d1}^{(1)}(\t)&= 
 \r(\t) \int_{\t^{--}/C_{1\dr}/C_{\d1 q}} \,c_{1\dr} \,c_{\d1 q}\,g^a_{0|1\dr}\,g_{0|\d1 q}\, g_{\d1 q|1\dr}\,e^{i\p_0+i\phi_{1\dr}+i\dphi_{\d1 q}}
 \\
&+  \r(\t)\int_{\t^{--}/C_{1\dr}} \,c_{1\dr} \,c_{\d1 q}(\t)\,g^a_{0|1\dr}\,R_{0|\d1 q}\, g_{1\dr|\d1 q}(\a_{1\dr}-\t)\,e^{i\p_0+i\phi_{1\dr}+i\dphi_{\d1 q}(\t)}
\\
&+ \r(\t) \int_{\t^{--}/C_{1\dr}} \,c_{1\dr} \,c_{\d1 q}(\a_{1\dr})\,g^a_{0|1\dr}\,g_{0|\d1 q}(\t-\a_{1\dr})\, R_{1\dr|\d1 q}\,e^{i\p_0+i\phi_{1\dr}+i\dphi_{\d1 q}(\a_{1\dr})}\,.
\eal
Here and in what follows  
the replacement $g_{1\dr|\d1 q}\to g_{\d1 q|1\dr}$ is done to stress that the contour $C_{\d1 q}$ is below $C_{1\dr}$.
Finally, $\dchi_\d1^- Z_{2\d1}^{(2)}$ is given by 
\bal\la{z2d24a}
 \dchi_\d1^- Z_{2\d1}^{(2)}(\t)&=  \r(\t)\int_{\t^{--}/C_{\d1 q}} \,c_{1\dr}(\t) \,c_{\d1 q}\,R_{0|1\dr}\,g_{0|\d1 q}(\t- \a_{\d1 q})\, g_{\d1 q|1\dr}(\a_{\d1 q}-\t)\,e^{i\p_0+i\phi_{1\dr}(\t)+i\dphi_{\d1 q}}\\
 &+  \r(\t)\,c_{1\dr}(\t) \,c_{\d1 q}(\t)\,R_{0|1\dr}\,R_{0|\d1 q}\,e^{i\p_0+i\phi_{1\dr}(\t)+i\dphi_{\d1 q}(\t)}\\
 &+  \r(\t)\,c_{1\dr}(\t) \,c_{\d1 q}(\t)\,R_{0|1\dr}\, R_{1\dr|\d1 q}\,e^{i\p_0+i\phi_{1\dr}(\t)+i\dphi_{\d1 q}(\t)}
 \,,
 \eal
where in the last two terms one used that $R_{0|1 \dr}=\de_{\dr\d1}R_{0|1\d1}$ and $R_{0|\d1 q} = \de_{q1}R_{0|\d1 1}$, and $g_{0|1\dt}=g_{\d1 1|1\d1}=1$.
Summing up the terms, and taking into account that $R_{0|1 \dr}=\de_{\dr\d1}R_{0|1\d1}$ and $R_{0|\d1 r}=\de_{r1}R_{0|\d1 1}$, one gets
\eqref{Z2d2}.

%%%%%%%%%%%%%%%%%%%%%%%%%%%%%%
\section{Traces of vertex operators}\la{traces}
\subsection*{General formula}
%%%%%%%%%%%%%%%%%%%%%%%%%%%%%
To compute traces of products of vertex operators 
defined as
\be
V(\t)=
 :\exp(i \phi(\t)):\,,
 \ee
where $\phi(\t)$ is a linear combination of the $N^2-1$ elementary oscillators $a_{mn}(t)$
\be
\phi(\t) = \int^\infty_{-\infty}{dt\over i t}\,\P_{A}(t)\,a_{A}(t)e^{i\theta t}= \int^\infty_{0} {dt\over i t}\,\bar\a_{A}(t)\,a_{A}(t) - \int^\infty_{0} {dt\over i t}\,a_{A}^\dagger(t)\b_{A}(t)\,,
\ee
%and 
\be
\bar\a_{A}(t)=\P_{A}(t)e^{i\theta t}\,,\quad \b_{A}(t)= \P_{A}(-t)e^{-i\theta t}\,,\quad a_{A}^\dagger(t) \equiv a_{A}(-t)\,,
\ee
it is sufficient to know how to compute 
\be\la{trV}
\mbox{Tr}_F\left(\exp(2\pi i K) V\right)\,,\quad K = iH=i\int^{\infty}_{0}dt \, \sum_{A}a_{A}^\dagger(t) a_{A}(t)\,,~~~~
\ee
where $F$ is the Fock space where $a_{A}(t)$ act, and 
\bal
V=e^{i\p}\,,\quad \phi = \int^\infty_{0} {dt\over i t}\,\bar\a_{A}(t)\,a_{A}(t) - \int^\infty_{0} {dt\over i t}\,a_{A}^\dagger(t)\b_{A}(t)\,.
\eal
The formula takes the following form 
\be\la{trVf}
{\mbox{Tr}_F\left(\exp(2\pi i K) V\right) \ov \mbox{Tr}_F\left(\exp(2\pi i K)\right) }= \exp\left(\int_0^\infty\, {dt\ov t}\,{\bar\alpha_{A}(t)\beta_{A}(t)\ov 1-e^{2\pi t}}\right)\,,
\ee
and its derivation can be found in e.g. \cite{BF13}.
This formula agrees with the prescription in \cite{FL}.
To show this let's consider
\be\la{trVV0}
\mbox{Tr}_F\left(\exp(2\pi i K) V_2V_1\right) \,,
\ee
where
\be
V_k=
 :\exp(i \phi_k):\,,\quad
\phi_k = \int^\infty_{0}{dt\ov it} \,\bar\a_{A}^{(k)}(t)\,a_A(t) - \int^\infty_{0}{dt\ov it}\,a_A^\dagger(t)\b_A^{(k)}(t)\,.
\ee
Then one gets
\be
V_2V_1 = g_{12}:V_2V_1:\,,\quad g_{12} =\exp\left( - \int_0^\infty\, {dt\ov t}\, \bar\alpha_A^{(2)}(t)\beta_A^{(1)}(t)\right)\,,
\ee
and
\be\la{trVV1}
{\mbox{Tr}_F\left(\exp(2\pi i K) V_2V_1\right) \ov \mbox{Tr}_F\left(\exp(2\pi i K)\right) }= \exp\left(  \int_0^\infty\, {dt\ov t}\, \left(-\bar\alpha_A^{(2)}(t)\beta_A^{(1)}(t) +{\bar\alpha_A(t)\beta_A(t)\ov 1-e^{2\pi t}}\right)\right)\,,
\ee
where
\be
\bar\alpha_A=\bar\alpha_A^{(1)}+\bar\alpha_A^{(2)}\,,\quad \beta_A(t)=\beta_A^{(1)}+\beta_A^{(2)}\,.
\ee
Thus,
\be\la{trVV2}
{\mbox{Tr}_F\left(\exp(2\pi i K) V_2V_1\right) \ov \mbox{Tr}_F\left(\exp(2\pi i K)\right) }= C_1 C_2 G_{12}\,,
\ee
where 
\be
C_k={\mbox{Tr}_F\left(\exp(2\pi i K) V_k\right) \ov \mbox{Tr}_F\left(\exp(2\pi i K)\right) }\,,
\ee
and 
\be\la{G12}
G_{12}= \exp\left(-  \int_0^\infty\, {dt\ov t}\, \left({\bar\alpha_A^{(1)}(t)\beta_A^{(2)}(t)\ov e^{2\pi t}-1}+{\bar\alpha_A^{(2)}(t)\beta_A^{(1)}(t)\ov 1-e^{-2\pi t}}\right)\right)\,,
\ee
Introducing 
\bea\begin{aligned}
\langle\langle a_A(t)a_B(t')\rangle\rangle&={t\de_{AB} \ov 1- e^{-2 \pi t}}\delta(t+t')\, ,
\end{aligned}\eea
one finds
\bea\begin{aligned}
\langle\langle\phi_2\phi_1\rangle\rangle&=-\log G_{12}\, .
\end{aligned}\eea
The generalisation to the product of $n$ vertex operators
\be
U_j(\t) =\, :e^{i\p_j(\t)}:\, =\, e^{i\p_j^+(\t)}\, e^{i\p_j^-(\t)}\,,\quad 
\ee
 is straightforward, and one gets 
\be
\langle\langle U_n(\t_n)\cdots U_1(\t_1) \rangle\rangle =\prod_{j=1}^nC_{U_j}\prod_{k<j}G_{U_kU_j}(\t_k-\t_j)\,, 
\ee
where
\be
C_{U_j}=\langle\langle U_j(\t_j)\rangle\rangle = \exp\big(-\langle\langle \p^-_j(0)\p^+_j(0)\rangle\rangle  \big)\,,
\ee
\be
G_{U_kU_j}(\t_k-\t_j) = \exp\big(-\langle\langle \p_j(\t_j)\p_k(\t_k)\rangle\rangle  \big)\,.
\ee

%%%%%%%%%%%%%%%%%%%%%%%%%%%
\subsection*{Traces of single $V$'s}
The traces of the vertex operators of the fields $\p_0$, $\p_{(r)}$, $\p_{k\dr}$, $\p_{\dk r}$ are given by
\be
C_\mu={\mbox{Tr}_F\left(e^{2\pi i K} V_\mu\right) \ov \mbox{Tr}_F\left(e^{2\pi i K} \right) }= \exp\left(\int_0^\infty\, {dt\ov t}\,{\P_{\mu,A}^{-}(t)\P_{\mu,A}^{+}(-t)\ov 1-e^{2\pi t}}\right)= \exp\left(\int_0^\infty\, {dt\ov t}\,{f_{\mu|\mu}(t)\ov 1-e^{2\pi t}}\right)\,,
\ee
where 
\bal
f_{\nu|\mu}(t)\equiv \P_{\mu,A}^{-}(t)\P_{\nu,A}^{+}(-t)\,,\quad t>0\,.
\eal
The functions $f_{\mu|\mu}(t)$ are given by
\bal
f_{0|0}(t) &=\Phi^-_{0,mn}(t)\Phi^+_{0,mn}(-t)= 2{\sinh {1\ov N}\pi t\,\sinh {N-1\ov N}\pi t\ov\sinh \pi t}\,,
\eal
\bal
f_{(r)|(r)}(t) &=\Phi^-_{(r),mn}(t)\Phi^+_{(r),mn}(-t)= 2{\sinh { r\ov N}\pi t\,\sinh {N-r\ov N}\pi t\ov\sinh \pi t}\,,
\eal
\bal
f_{k\dr|k\dr}(t) &=\Phi^-_{k\dr,mn}(t)\Phi^+_{k\dr,mn}(-t)= 1-e^{-2\pi t/N}\,,
\eal
\bal
f_{\dk r|\dk r}(t) &=\Phi^-_{\dk r,mn}(t)\Phi^+_{\dk r,mn}(-t)= 1-e^{-2\pi t/N}\,.
\eal
Notice that all these functions asymptote to 1 at large $t$, and therefore the integrals are well-defined, and can be computed explicitly by using formulae from appendix B of \cite{BF13}.

%%%%%%%%%%%%%%%%%%%%%%%%%%%
\subsection*{Traces of $V_\mu V_\nu$ and functions $G_{\mu\nu}$}

To compute the traces one uses \eqref{trVV2}, and 
\eqref{G12} which takes the form
\be
G_{\nu| \mu}(\b_2-\b_1)=\exp\big(-\langle\langle \p_\mu(\b_1) \p_\nu(\b_2)\rangle\rangle\big)\, ,
\ee
and therefore
\bal
G_{\nu| \mu}(\b)&=\exp\left(-\int_0^\infty\, {dt\ov t}\, \left({f_{\nu|\mu}(t)e^{-i\b t}\ov 1-e^{-2\pi t}}+{f_{\mu|\nu}(t)e^{i\b t}\ov e^{2\pi t}-1}\right)\right)
\, .
\eal
These satisfy the relations
\be
G_{\mu| \nu}(\b-2\pi i)=G_{\nu| \mu}(-\b)\, ,\quad S_{\mu| \nu}(\b)={G_{\nu| \mu}(-\b)\ov G_{\mu| \nu}(\b)}={G_{\mu| \nu}(\b-2\pi i)\ov G_{\mu| \nu}(\b)}\, ,
\ee
which are necessary to satisfy the form factors axioms.

If $f_{\nu|\mu}(t)=f_{\mu|\nu}(t)$, in particular, for $\nu=\mu$, one gets the familiar representation
\bal
G_{\nu|\mu}(\b)=G_{\mu|\nu}(\b)=
&=\exp\left(-\int_0^\infty\, {dt\ov t}\, {f_{\nu|\mu}(t)\cos(\b+i\pi) t\ov \sinh{\pi t}}\right)\,,\quad f_{\nu|\mu}(t)=f_{\mu|\nu}(t)
\, .
\eal
Thus, taking into account that $f_{(r)(q)}(t)=f_{(q)(r)}(t)$, one gets
the following representations for $G_{(r)|(q)}$ (note that $G_{0(q)}\equiv G_{(1)|(q)}$)
\bal\la{GNm11}
G_{(r)|(q)}(\b)
&=\exp\left(-2\int_0^\infty\, {dt\ov t}\, {{\sinh {q\ov N}\pi t\,\sinh {N-r\ov N}\pi t}\cos(\b+i\pi) t\ov \sinh^2{\pi t}}\right)
\, ,\quad r\ge q\,,
\eal
which is well-defined for 
\be
 -2 \pi-{|r-q|\ov N}\pi  <{\rm Im}(\beta )<{|r-q|\ov N}\pi\,.
\ee
The functions $G_{k\dr|k\dr}$ and $G_{\dk r|\dk r}$ are equal to each other 
\bal
G_{k\dr|k\dr}(\b)
&=G_{\dk r|\dk r}(\b)=\exp\left(-\int_0^\infty\, {dt\ov t}\, {(1-e^{-2\pi t/ N})\cos(\b+i\pi) t\ov \sinh{\pi t}}\right)
\, ,
\eal
and computing them by using formulae from appendix B of \cite{BF13}, one gets
\bal
G_{k\dr|k\dr}(\b)
&=e^{{2\g/ N}}(2\pi)^{2/N}{i\ov\pi}\G\big({1\ov N}+{i\b\ov2\pi}\big)\G\big(1+{1\ov N}-{i\b\ov2\pi}\big){\sinh{\b\ov 2}}
\, .
\eal
The function has poles at 
\bal
\b={2\pi\ov N}i+2\pi i\, m\ {\rm \ and}\ \  \b=-{2\pi(N+1)\ov N}i-2\pi i\, m\,,\ \ m=0,1,2,\ldots\,,
\eal
and the integration contour in $\b$ should run below the poles $\b=2\pi i/N+2\pi i\, m$ but above $ \b=-{2\pi(N+1)\ov N}i-2\pi i\, m$ because the only pole of the function $g_{k\dr|k\dr}(\b)$ is at $\b={2\pi\ov N}i$.

One also finds
\bal
f_{0|1\d1}(t)&=f_{0|\d1 1}(t)=e^{-2\pi t/ N}-1\,,\quad f_{1\d1|0}(t)=f_{\d1 1|0}(t)=0\,,\\
G_{0|1\d1}(\b)
&=G_{0|\d1 1}(\b)=\exp\left(-\int_0^\infty\, {dt\ov t}\, {(e^{-2\pi t/ N}-1)e^{-i\b t}\ov 1-e^{-2\pi t}}\right)\\
&=e^{-{\g/ N}}(2\pi)^{-1/N}{\G\big({i\b\ov2\pi}\big)\ov \G\big({1\ov N}+{i\b\ov2\pi}\big)}
\, ,
\eal
$G_{0|1\d1}(\b)$ has poles at $\b=2\pi i\,m$, and in the product of Green's functions containing $G_{0|1\d1}(\t-\a_{1\d1})$ the integration contour $C_{\a_{1\d1}}$ runs above 
the poles at $\a_{1\d1}=\t-2\pi i\,m$. 
\bal
G_{1\d1|0}(\b)
&=G_{\d1 1|0}(\b)=\exp\left(-\int_0^\infty\, {dt\ov t}\, {(e^{-2\pi t/ N}-1)e^{i\b t}\ov e^{2\pi t}-1}\right)\\
&=e^{-{\g/ N}}(2\pi)^{-1/N}{\G\big(1-{i\b\ov2\pi}\big)\ov \G\big(1+{1\ov N}-{i\b\ov2\pi}\big)}
\, .
\eal
$G_{1\d1|0}(\b)$ has poles at $\b=-2\pi i\,m\,,\ m=1,2,\ldots$, and in the product of Green's functions containing $G_{1\d1|0}(\a_{1\d1}-\t)$ the integration contour $C_{\a_{1\d1}}$ runs above 
all these poles at $\a_{1\d1}=\t-2\pi i\,m$. Then
\bal
G_{0|1\d1}^a(\b)
&=G_{0|1\d1}(\b)-G_{1\d1|0}(-\b)={e^{-{\g/ N}}(2\pi)^{-1/N}\G\big({i\b\ov2\pi}\big)\ov N\,\G\big(1+{1\ov N}+{i\b\ov2\pi}\big)}
\, ,
\eal
and the integration contour $C_{\a_{1\d1}}$ runs above 
its poles at $\a_{1\d1}=\t-2\pi i\,m$. 
 
\bal
f_{0|1\dr}(t)&=f_{0|\d1 r}(t)=0\,,\quad f_{1\dr|0}(t)=f_{\d1 r|0}(t)=1-e^{2\pi t/ N}\,,\quad r,\dr\ge 2\,,\\
G_{1\dr|0}(\b)
&=G_{\d1 r|0}(\b)=\exp\left(-\int_0^\infty\, {dt\ov t}\, {(1-e^{2\pi t/ N})e^{-i\b t}\ov 1-e^{-2\pi t}}\right)\\
&=e^{-{\g/ N}}(2\pi)^{-1/N}{\G\big(-{1\ov N}+{i\b\ov2\pi}\big)\ov \G\big({i\b\ov2\pi}\big)}
\, ,
\eal
$G_{1\dr|0}(\b)$ has poles at $\b=-2\pi i/N+2\pi i\,m$, and in the product of Green's functions containing $G_{1\dr|0}(\a_{1\dr}-\t)$ the integration contour $C_{\a_{1\dr}}$ runs below 
all these poles at $\a_{1\dr}=\t-2\pi i/N+2\pi i\,m$. 
\bal
G_{0|1\dr}(\b)
&=G_{0|\d1 r}(\b)=\exp\left(-\int_0^\infty\, {dt\ov t}\, {(1-e^{2\pi t/ N})e^{i\b t}\ov e^{2\pi t}-1}\right)\\
&=e^{-{\g/ N}}(2\pi)^{-1/N}{ \G\big(1-{1\ov N}-{i\b\ov2\pi}\big)\ov\G\big(1-{i\b\ov2\pi}\big)}
\, ,
\eal
$G_{0|1\dr}(\b)$ has poles at $\b=-2\pi i(N-1)/N-2\pi i\,m$, and in the product of Green's functions containing $G_{0|1\dr}(\t-\a_{1\dr})$ the integration contour $C_{\a_{1\dr}}$ runs below 
all these poles at $\a_{1\dr}=\t+2\pi i(N-1)/N+2\pi i\,m$. 
Then
\bal
G_{0|1\dr}^a(\b)
&=G_{0|1\dr}(\b)-G_{1\dr|0}(-\b)={e^{-{\g/ N}}(2\pi)^{-1/N}\G\big(-{1\ov N}-{i\b\ov2\pi}\big)\ov N\,\G\big(1-{i\b\ov2\pi}\big)}
\, ,
\eal
and the integration contour $C_{\a_{1\dr}}$ in $G^a_{0|1\dr}(\t-\a_{1\dr})$ runs below its poles at $\a_{1\dr}=\t-2\pi i/N+2\pi i\,m$.

\bal
f_{k\dr|k\dq}(t)&=f_{\dk r|\dk q}(t)=0\,,\quad f_{k\dq|k\dr}(t)=f_{\dk q|\dk r}(t)=e^{2\pi t/ N}-e^{-2\pi t/ N}\,,\quad r<q,\dr<\dq\,,\\
G_{k\dr|k\dq}(\b)&=G_{\dk r|\dk q}(\b)
=\exp\left(-\int_0^\infty\, {dt\ov t}\, {(e^{2\pi t/ N}-e^{-2\pi t/ N})e^{i\b t}\ov e^{2\pi t}-1}\right)\\
&=e^{-{\g/ N}}(2\pi)^{-1/N}{ \G\big(1-{1\ov N}-{i\b\ov2\pi}\big)\ov\G\big(1-{i\b\ov2\pi}\big)}
\, ,
\eal
$G_{k\dr|k\dq}(\b)$, $\dr<\dq$ has poles at $\b=-2\pi i(N-1)/N-2\pi i\,m$, and in the product of Green's functions containing $G_{k\dr|k\dq}(\a_{k\dr}-\a_{k\dq})$ the integration contour $C_{\a_{k\dq}}$ runs below 
all these poles at $\a_{k\dq}=\a_{k\dr}+2\pi i(N-1)/N+2\pi i\,m$. 
\bal
G_{k\dq|k\dr}(\b)&=G_{\dk q|\dk r}(\b)
=\exp\left(-\int_0^\infty\, {dt\ov t}\, {(e^{2\pi t/ N}-e^{-2\pi t/ N})e^{-i\b t}\ov 1-e^{-2\pi t}}\right)\\
&=e^{{2\g/ N}}(2\pi)^{2/N}{\G\big({1\ov N}+{i\b\ov2\pi}\big)\ov \G\big(-{1\ov N}+{i\b\ov2\pi}\big)}
\, ,
\eal
$G_{k\dq|k\dr}(\b)$, $\dr<\dq$ has poles at $\b=2\pi i/N+2\pi i\,m$, and in the product of Green's functions containing $G_{k\dq|k\dr}(\a_{k\dq}-\a_{k\dr})$ the integration contour $C_{\a_{k\dq}}$ runs below 
all these poles at $\a_{k\dq}=\a_{k\dr}+2\pi i/N+2\pi i\,m$. 
\bal
\left\{
\begin{array}{c}
 G_{k\dr|k+1,\dq}(\b)=G_{\dk r|\dk+\d1,q}(\b)=G_{0|1\d1}(\b)\,, \\
 G_{k+1,\dq|k\dr}(\b)=G_{\dk+\d1,q|\dk r}(\b)=G_{1\d1|0}(\b)\,,     
\end{array}
\right.
 \qquad  \dr\ge \dq\,,\ r\ge q\,,
\eal
\bal
\left\{
\begin{array}{c}
 G_{k\dr|k+1,\dq}(\b)=G_{\dk r|\dk+\d1,q}(\b)=G_{0|1\dt}(\b)\,, \\
 G_{k+1,\dq|k\dr}(\b)=G_{\dk+\d1,q|\dk r}(\b)=G_{1\dt|0}(\b)\,,     
\end{array}
\right.
 \qquad  \dr< \dq\,,\ r< q\,,
\eal
\bal
f_{k\dr|\dr, k+1}(t)&=f_{\dr k|k, \dr+\d1}(t)=e^{-2\pi t/ N}-1\,,\quad 
f_{\dr, k+1|k\dr}(t)=f_{k, \dr+\d1|\dr k}(t)=e^{2\pi t/ N}-1\,,
\\
G_{k\dr|\dr, k+1}(\b)&=G_{\dr k|k, \dr+\d1}(\b)=\exp\left(-\int_0^\infty\, {dt\ov t}\, \left({(e^{-2\pi t/ N}-1)e^{-i\b t}\ov 1-e^{-2\pi t}}+{(e^{2\pi t/ N}-1)e^{i\b t}\ov e^{2\pi t}-1}\right)\right)\\
&={G_{0|1\d1}(\b)\ov G_{0|1\dt}(\b)}={\sinh \left(\frac{\beta }{2}-\frac{\pi i }{N}\right)\ov \sinh\left(\frac{\beta }{2}\right)}\,,
\\ 
G_{\dr, k+1|k\dr}(\b)&=G_{k, \dr+\d1|\dr k}(\b)=\exp\left(-\int_0^\infty\, {dt\ov t}\, \left({(e^{2\pi t/ N}-1)e^{-i\b t}\ov 1-e^{-2\pi t}}+{(e^{-2\pi t/ N}-1)e^{i\b t}\ov e^{2\pi t}-1}\right)\right)\\
&={\sinh \left(\frac{\beta }{2}+\frac{\pi i }{N}\right)\ov \sinh\left(\frac{\beta }{2}\right)}\,.
\eal
These functions have poles at $\b=2\pi i m$, $m\in \bZ$. The contour should run below the poles with nonpositive imaginary part but above the poles with negative imaginary part. 
\bal
f_{r\dq|(r)}(t )&=f_{\dr q|(r)}(t )=e^{(r-1)\pi t/N}-e^{(r+1)\pi t/N}\,,\quad f_{(r)|r\dq}(t )=f_{(r)|\dr q}(t )=0\,,\quad \dq > \dr\,,
\\
G_{r\dq|(r)}(\b)&=G_{\dr q|(r)}(\b)
=\exp\left(-\int_0^\infty\, {dt\ov t}\, {(e^{(r-1)\pi t/N}-e^{(r+1)\pi t/N})e^{-i\b t}\ov 1-e^{-2\pi t}}\right)\\
&=e^{-{\g/ N}}(2\pi)^{-1/N}{\G\big(-{r+1\ov 2N}+{i\b\ov2\pi}\big)\ov \G\big(-{r-1\ov 2N}+{i\b\ov2\pi}\big)}
\, ,
\eal
$G_{r\dq|(r)}(\b)$, $\dq > \dr$ has poles at $\b=-\pi i(r+1)/N+2\pi i\,m$, and in the product of Green's functions containing $G_{r\dq|(r)}(\a_{r\dq}-\t)$ the integration contour $C_{\a_{r\dq}}$ runs below 
all these poles at $\a_{r\dq}=\t-\pi i(r+1)/N+2\pi i\,m$. 
\bal
G_{(r)|r\dq}(\b)&=G_{(r)|\dr q}(\b)
=\exp\left(-\int_0^\infty\, {dt\ov t}\, {(e^{(r-1)\pi t/N}-e^{(r+1)\pi t/N})e^{i\b t}\ov e^{2\pi t}-1}\right)\\
&=e^{-{\g/ N}}(2\pi)^{-1/N}{\G\big(1-{r+1\ov 2N}-{i\b\ov2\pi}\big)\ov \G\big(1-{r-1\ov 2N}-{i\b\ov2\pi}\big)}
\, ,
\eal
$G_{(r)|r\dq}(\b)$, $\dq > \dr$ has poles at $\b=\pi i(r+1)/N-2\pi i\,m$, $m=1,2,\ldots$, and in the product of Green's functions containing $G_{(r)|r\dq}(\t-\a_{r\dq})$ the integration contour $C_{\a_{r\dq}}$ runs below 
all these poles at $\a_{r\dq}=\t-\pi i(r+1)/N+2\pi i\,m$. 
\bal
f_{r\dq|(r)}(t )&=f_{\dr q|(r)}(t )=0\,,\quad f_{(r)|r\dq}(t )=f_{(r)|\dr q}(t )=e^{-(r+1)\pi t/N}-e^{-(r-1)\pi t/N}\,,\quad \dq\le \dr\,,\\
G_{r\dq|(r)}(\b)&=G_{\dr q|(r)}(\b)
=\exp\left(-\int_0^\infty\, {dt\ov t}\, {(e^{-(r+1)\pi t/N}-e^{-(r-1)\pi t/N})e^{i\b t}\ov e^{2\pi t}-1}\right)\\
&=e^{-{\g/ N}}(2\pi)^{-1/N}{\G\big(1+{r-1\ov 2N}-{i\b\ov2\pi}\big)\ov \G\big(1+{r+1\ov 2N}-{i\b\ov2\pi}\big)}\, ,
\eal
$G_{r\dq|(r)}(\b)$, $\dq \le \dr$ has poles at $\b=-\pi i(r-1)/N-2\pi i\,m$, $m=1,2,\ldots$, and in the product of Green's functions containing $G_{r\dq|(r)}(\a_{r\dq}-\t)$ the integration contour $C_{\a_{r\dq}}$ runs above 
all these poles at $\a_{r\dq}=\t-\pi i(r-1)/N-2\pi i\,m$. 
\bal
G_{(r)|r\dq}(\b)&=G_{(r)|\dr q}(\b)
=\exp\left(-\int_0^\infty\, {dt\ov t}\, {(e^{-(r+1)\pi t/N}-e^{-(r-1)\pi t/N})e^{-i\b t}\ov 1-e^{-2\pi t}}\right)\\
&=e^{-{\g/ N}}(2\pi)^{-1/N}{\G\big({r-1\ov 2N}+{i\b\ov2\pi}\big)\ov \G\big({r+1\ov 2N}+{i\b\ov2\pi}\big)}
\, ,
\eal
$G_{(r)|r\dq}(\b)$, $\dq \le \dr$ has poles at $\b=\pi i(r-1)/N+2\pi i\,m$, and in the product of Green's functions containing $G_{(r)|r\dq}(\t-\a_{r\dq})$ the integration contour $C_{\a_{r\dq}}$ runs above 
all these poles at $\a_{r\dq}=\t-\pi i(r-1)/N-2\pi i\,m$.

%%%%%%%%%%%%%%%%%%%%%%%%%%%%%%
%%%%%%%%%%%%%%%%%%%%%%%%%%%%%%%%%%%%%%%%%%%%%%%%%

\end{document}